\RequirePackage{fix-cm}
\documentclass{svjour3-HAL}                     
\usepackage{verbatim}
\usepackage{stackengine}
\usepackage{amssymb}
\usepackage{amsmath}
\usepackage{graphicx}
\usepackage{comment}
\usepackage[noend]{algorithmic}

\usepackage{amsmath}
\usepackage{wasysym}
\usepackage{booktabs}

\usepackage[ruled]{algorithm}
\usepackage{algorithmic}

\usepackage{url}
\urldef{\mailsb}\path|nicolas.spyratos@lri.fr|

\usepackage{todonotes}
\newcommand{\comns}[2]{\todo[inline,color=red,caption={}]{\textbf{#1}: #2}}
\newcommand{\comm}[2]{\todo[inline,color=green,caption={}]{\textbf{#1}: #2}}

\begin{document}

\title{The Context Model: A Graph Database Model}



%
%

\author{Nicolas Spyratos}
\institute{LISN Laboratory - University Paris-Saclay, CNRS, INRIA, France\\
Adjunct Scienist, FORTH Institute of Computer Science,  Greece\\
\mailsb\\~\\
{\bf Acknowledgment:} Work conducted while the author was visiting with the HCI Group at FORTH Institute of Computer Science, Greece (https://www.ics.forth.gr/)}
\date{}

\maketitle

\begin{abstract}

We propose a novel database model whose basic structure is a labeled, directed, acyclic graph with a single root, in which the nodes represent the data sets 
of an  application and the edges represent functional relationships among the data sets. We call such a graph an {\em application context} or simply {\em context}. The query language of a context consists of two types of queries, traversal queries and analytic queries. Both types of queries are defined using a simple functional algebra whose operations are functional restriction, composition of functions, pairing of functions and Cartesian product of sets. Roughly speaking, traversal queries parallel relational algebra queries, whereas analytic queries parallel SQL Group-by queries. In other words, in our model, traversal queries {\em and} analytic queries, are both defined within the {\em same} formal framework - in contrast to the relational model, where analytic queries are defined outside the relational algebra. Therefore a distinctive feature of our model is that it supports data management {\em and} data analytics within the {\em same} formal framework.  

\noindent We demonstrate the expressive power of our model by showing: (a) how a relational database can be defined as a view over a context, with the context playing the role of an underlying semantic layer; (b) how an analytic query over a context can be rewritten at two orthogonal levels: at the level of the traversal queries that do the grouping and measuring, and at the level of the analytic query itself; and (c) how a context can be used as a user-friendly interface for querying relations and analysing relational data. 

\end{abstract}


\section{Introduction}

The basic idea underlying this work is that the data sets of an application and their relationships can be seen as a labeled, directed, acyclic graph with a single root, in which the nodes are the data sets and in which each edge is a function from its source node to its target node. We call such a graph an {\em application context} or simply a {\em context}. 


The concept of context was first introduced in \cite{DBLP:conf/fqas/Spyratos06}\cite{SpyratosS18} as a means for defining analytic queries, in the abstract, then translating them as queries in an underlying query evaluation mechanism such as SQL, MapReduce or SPARQL. In this paper we build upon these earlier works to propose a full fledged data model supporting both, data management {\em and} data analytics within the {\em same} formal framework.


To introduce the concept of context we borrow the example of \cite{SpyratosS18} that we shall use as our running example. Suppose $Inv$ is the set of all delivery invoices, say over a year, in a distribution center (e.g. Walmart) which delivers products of various types in a number of branches. A delivery invoice has an identifier (e.g. an integer) and shows the date of delivery, the branch in which the delivery took place, the type of product delivered (e.g. $Coca Light$) and the quantity (i.e. the number of units delivered of that type of product). There is a separate invoice for each type of product delivered, and the data on all invoices during the year is stored in a database for analysis and planning purposes.  

\begin{figure}\label{CxtDef}
{
\begin{center}
\includegraphics[width=350px,keepaspectratio]{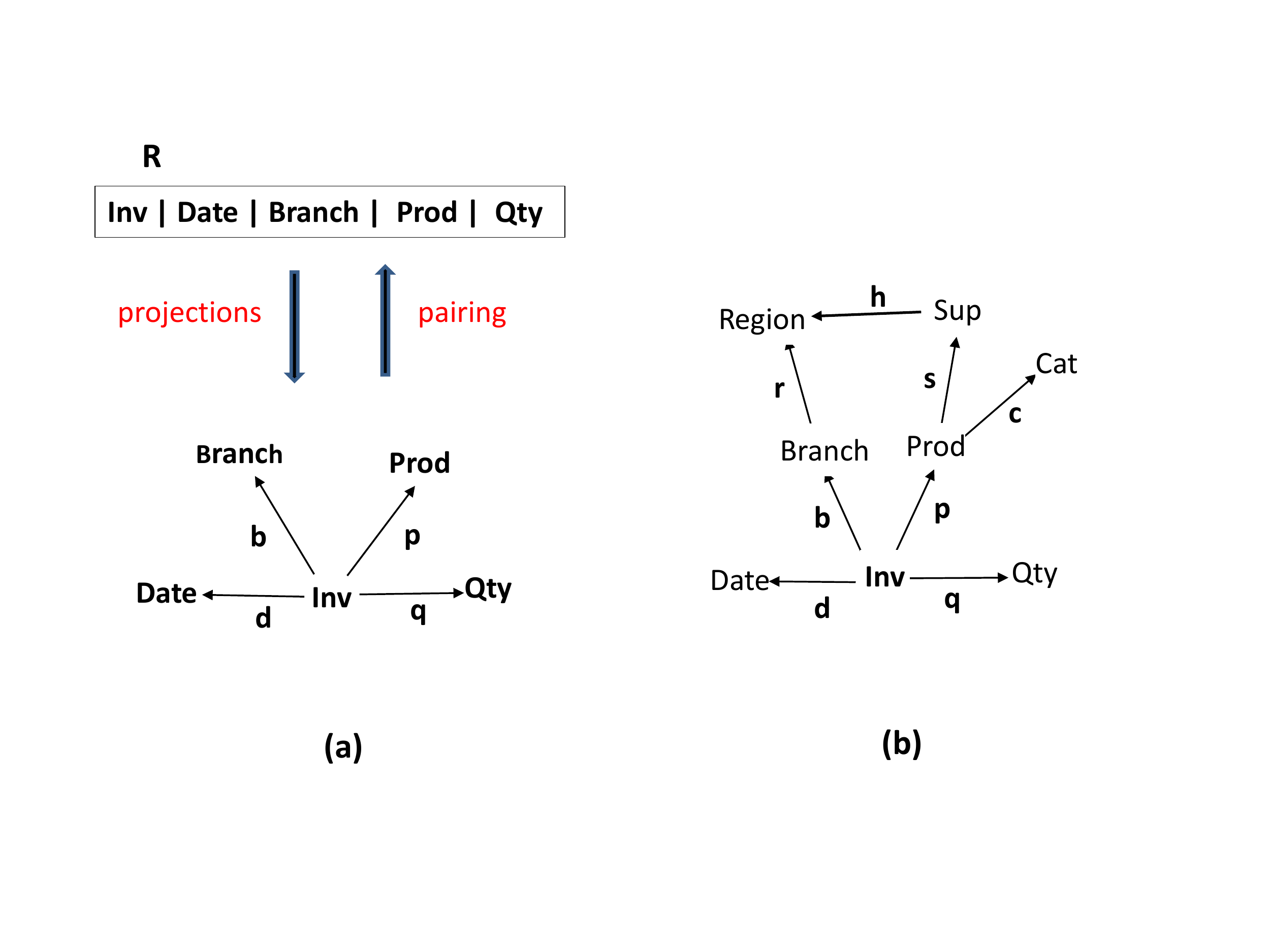}
\caption{Example of context definition\label{fig-1}}
\end{center}
}
\end{figure}

Conceptually, the information provided by each invoice would most likely be represented as a tuple of a relation $R$ with the following attributes: Invoice number ($Inv$), $Date$, $Branch$, Product ($Prod$) and Quantity ($Qty$), with Invoice number as the key. Now, as $Inv$ is the key, we have the following key dependencies:  
\begin{center}
$d: Inv \to Date, ~~b: Inv \to Branch, ~~p: Inv \to Prod, ~~q: Inv \to Qty$     
\end{center}
\noindent These dependencies form a graph $\mathcal C$ as shown in Figure~\ref{CxtDef}(a) (actually a tree in this case). 

Think now of a mapping $\delta$ that associates the nodes and edges of $\mathcal C$ with `values' using projections over $R$ as follows: \\

\noindent {\bf Nodes}: $\delta(Inv)= \pi_{Inv}(R)$, ~~$\delta(Date)= \pi_{Date}(R)$, ~~$\delta(Branch)= \pi_{Branch}(R)$, $\delta(Prod)= \pi_{Prod}(R)$, ~~$\delta(Qty)= \pi_{Qty}(R)$

\noindent {\bf Edges}: $\delta(d)=\pi_{Inv, Date}(R)$, ~~$\delta(b)=\pi_{Inv, Branch}(R)$, ~~$\delta(p)=\pi_{Inv, Prod}(R)$, ~~$\delta(q)=\pi_{Inv, Qty}(R)$ \\ 

Clearly, if the only dependencies are key dependencies and $R$ is consistent then each of the associations $\delta(d)$, $\delta(b)$, $\delta(p)$ and $\delta(q)$ is a total function; and all these functions have the same domain of definition, namely the projection $\pi_{Inv}(R)$ of relation $R$ over its key $Inv$. 


Following this view, given an invoice number $i$ in $\delta(Inv)$, the function $\delta(d)$ returns a date $\delta(d)(i)$, the function $\delta(b)$ returns a branch $\delta(b)(i)$, the function $\delta(p)$ returns a product type $\delta(p)(i)$ and the function $\delta(q)$ returns a quantity $\delta(q)(i)$ (i.e. the number of units of product type $\delta(p)(i)$). Moreover by `pairing' these four functions we recover the relation $R$ that is: 
\begin{center}
 $R= \delta(d) \wedge \delta(b) \wedge \delta(p) \wedge \delta(q)$, where `$\wedge$' denotes the operation of pairing     
\end{center}

In this paper, given two functions $f: X \to Y$ and $g: X \to Z$ with $X$ as their common source, we call {\em pairing} of $f$ and $g$, denoted as $f \wedge g$, the function defined as follows:
\begin{center}
 $f \wedge g: X \to Y \times Z$ such that $(f \wedge g)(x)= (f(x), g(x))$ for all $x$ in $X$.    
\end{center}
Note that pairing works as a tuple constructor. Indeed, if we view the elements of $X$ as identifiers, then for each $x$ in $X$ the pairing constructs a tuple $(f(x), g(x))$ of the images of $x$ under the input functions; and this tuple is identified by $x$. In other words, the graph of the function $f \wedge g$ is a set of triples of the form $(x, f(x), g(x))$, therefore a relation over $\{X, Y, Z\}$ having $X$ as key and `satisfying' the functional dependencies $X \to Y$ and $X \to Z$. Clearly, the definition of pairing can be extended to more than two functions with the same source in a straightforward manner. As we shall see shortly the operation of pairing plays a fundamental role in our model.

Figure~\ref{CxtDef}(a) shows the one-one correspondence between consistent relations and contexts. As we see in this figure, we go from relations to contexts using projection and from contexts to relations using pairing.


 In this paper we propose a model in which contexts are treated as `first class citizens' in the sense that we study the concepts of context and database over a context in their own right (i.e. as a separate data model) 
and then we use the results of our study to gain more insights into some fundamental concepts of data management and data analytics. 

As another example of context, suppose that, apart from the relation $R$, we have three more relations defined as follows:

$R1(Branch, Region)$ with $Branch$ as key 

$R2(Prod, Sup, Cat)$ with $Prod$ as key 

$R3(Sup, Region)$ with $Sup$ as key

\smallskip\noindent
The relation $R1$ gives for each branch the region where the branch is located; the relation $R2$ gives for each product the supplier and category of that product; and the relation $R3$ gives for each supplier the region in which the supplier's seat is located. Following the same reasoning as for $R$, we can associate $R1$ with the context $\{Branch \to Region\}$, $R2$ with the context $\{Prod \to Sup, Prod \to Cat\}$ and $R3$ with the context $\{Sup \to Region\}$. These three contexts put together with the context associated with $R$ make up the context shown in Figure~\ref{CxtDef}(b).

Another way to look at the context shown in Figure~\ref{CxtDef}(b) is to think of it as an `evolution' of the context in Figure~\ref{CxtDef}(a) in the sense that we have added the region in which each branch is located; the category and the supplier of each product; and the region in which each supplier has its seat. In other words, we can model this application {\em independently} of its relational representation or of any other representation for that matter. In a way, our model is related to the entity-relationship model \cite{DBLP:journals/tods/Chen76}, in the sense that (a) it adopts the more natural view that the real world consists of entities (here, the nodes) and relationships (here, functional relationships) and (b) it achieves a high degree of data independence. 

Incidentally, note that the new edges added to the context of Figure~\ref{CxtDef}(a), namely $r$, $c$, $s$ and $h$ can be seen as `derived' edges in the sense that they can be computed from the existing ones. Indeed, the region of each branch can be computed from the address of the branch  (by Geo-localization); the region of each supplier can be computed from the address of the supplier's seat; and the category and supplier of each product can be `read off' the code bar of the product. 
Also note that the context of Figure~\ref{CxtDef}(b) has two `parallel paths' from $Inv$ to $Region$ (`parallel' in sense `same source and same target'). By the way, the context of Figure~\ref{CxtDef}(b) could be the context of a data warehouse with fact table $R$ and dimension tables $R1, R2, R3$. 

\begin{figure}
{
\begin{center}
\includegraphics[width=350px,keepaspectratio]{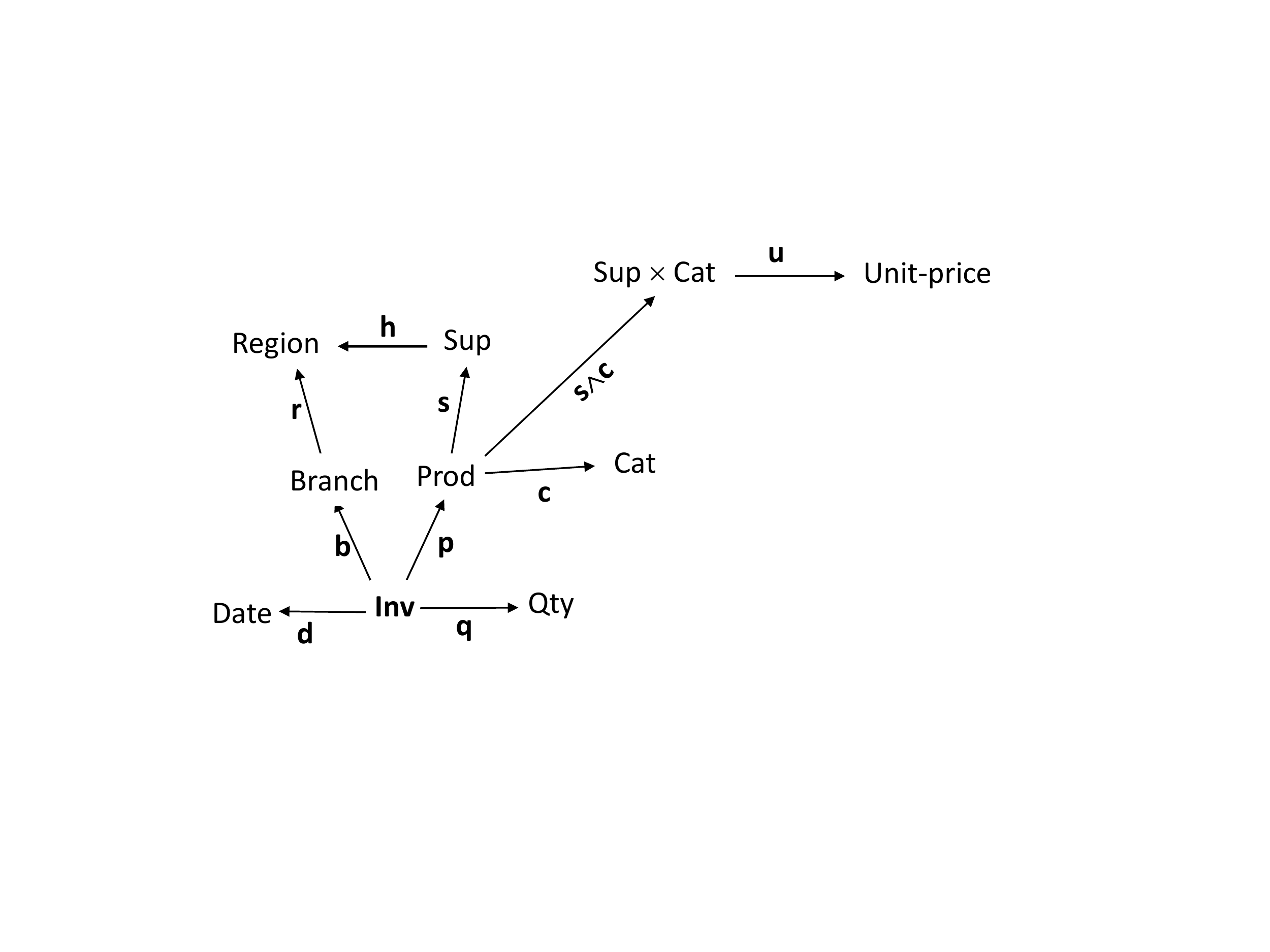}
\caption{Representing the dependency $\{Sup, Cat\} \to Unitprice$\label{Price}}
\end{center}
}
\end{figure}

As a last remark, consider the context of Figure~\ref{CxtDef}(b) and suppose that the price of a product is determined by the supplier {\em and} the category of the product. In the relational model this is expressed by defining a relational schema $R4(Cat, Prod, Unitprice)$ with the functional dependency $\{Sup, Cat\} \to Unitprice$. To express this dependency in our model we need to add the edges $s\wedge c: Prod \to Sup \times Cat$ and $u: Sup \times Cat \to Unitprice$ as shown in Figure~\ref{Price}.

It should be evident from our examples that the model that we propose here is a graph database model \cite{DBLP:reference/bdt/GutierrezHW19}. A graph database model is a schema-less model that uses nodes, relationships between nodes and key-value properties instead of tables to represent information. Therefore it is typically substantially faster for associative data sets. While other database models compute relationships expensively at query time, a graph database stores connections as first-class citizens, readily available for any `join-like' navigation operation. Any purposeful query can be easily answered via graph databases, as data is easily accessed using traversals. A traversal is how you query a graph, navigating from starting nodes to related nodes according to an algorithm \cite{DBLP:journals/corr/AnglesABHRV16}. As we shall see, the functional algebra that we use in our model allows to define traversal queries, as well as  analytic queries in which grouping and measuring is done by traversal queries. 


We demonstrate the expressive power of our model by showing: (a) how a relational database can be defined as a view over a context, with the context playing the role of an underlying semantic layer, (b) how an analytic query over a context can be rewritten at two orthogonal levels, namely at the level of the traversal queries that do the grouping and measuring for the analytic query, and at the level of the analytic query itself and (c) how a context can be used as a user-friendly interface for querying relations and analysing relational data. 

There is a substantial body of literature around the use of graphs in computer science, including several tools to support graph management by means of digital technology, as well as data models and query languages based on graphs. The reader is referred to \cite{DBLP:conf/sigmod/ArenasGS21}\cite{DBLP:conf/aib/Hogan22}\cite{DBLP:journals/csur/HoganBCdMGKGNNN21} for a comprehensive analysis of graph-based approaches to data and knowledge management, emphasizing the relation between graph databases and knowledge graphs,  and including an extensive bibliography. A rather detailed discussion on the power and limitations of graph databases can be found in \cite{DBLP:conf/cisim/Pokorny15}. Moreover, several graph database systems have appeared in recent years such as Neo4j Graph Database, ArangoDB, Amazon Neptune, Dgraph and a host of others\footnote{https://www.g2.com/categories/graph-databases (retrieved on April 26, 2023)}.

The remaining of the paper is organized as follows. In section~\ref{sec:Formal} we present the formal model, namely the basic concepts of context and database over a context; the functional algebra for defining traversal queries and analytic queries; and the concept of integrity constraint. In section~\ref{sec:TQ} we present the language of traversal queries over a context and their evaluation techniques, as well as the concept of view over a context. In section~\ref{sec:AQ} we present the language of analytic queries and their evaluation techniques, in particular their rewriting techniques. In section ~\ref{sec:Apps} we present applications of our model illustrating its expressive power. Finally, in section~\ref{sec: Conclusions}, we offer concluding remarks and discuss perspectives of our work.


\section{The formal model}\label{sec:Formal}

In this section we build upon the work of \cite{DBLP:conf/fqas/Spyratos06}\cite{SpyratosS18} to give the formal definition of context and of database over a context; we then present our functional algebra for manipulating the database and we define the integrity constraints that we use in our model. 


\subsection{The definition of context}

As we have seen informally in the introduction, a context is a labeled directed acyclic graph with a single root, in which each node represents a data set and each edge $f$ from a node $X$ to a node $Y$ represents a function from the data set $X$ to the data set $Y$. 

To introduce the formal definition of context we need some auxiliary concepts. Let $U$ be a finite set that we shall call the {\em universe}. A {\em node} over $U$ is either an element of $U$ ({\em simple node}) or the Cartesian product of a finite set of elements of $U$ ({\em composite node} or {\em product node}). For example, in Figure~\ref{Price}, $Branch$ or $Price$ are simple nodes, whereas $Sup \times Cat$ is a composite node. We assume that the data set represented by a simple node $A$ comes always from a given, fixed set of values associated with $A$, called the {\em domain} of $A$ and denoted as $dom(A)$; moreover, we assume that two simple nodes can have the same domain. As for a product node its domain is defined to be the product of the domains of its factors; for example, $dom(A \times B)= dom(A) \times dom(B)$. Although the domain of a node can be an infinite set, a node (whether simple or composite) always represents a {\em finite} set of values from its domain. 

Now, a context being a graph, it may contain cycles. However, we assume that a context is always an acyclic graph. Indeed, if a context has cycles then we can convert it into an acyclic graph by (a) considering that all nodes in a cycle are equivalent and (b) by coalescing all nodes in every cycle to a single node. To define the sense in which all nodes of a cycle are equivalent, we proceed as follows: (a) we assume that, for every node $X$, there is an `identity edge' $\iota_X: X \to X$ and (b) for all nodes $X$ and $Y$, we define $X \le Y$ if there is a path from $X$ to $Y$. Then the relation defined by: $X \equiv Y$ if $X \le Y$ and $Y \le X$ is an equivalence relation over the nodes of a cycle (as it is reflexive, symmetric and transitive). It follows that all nodes in a cycle are equivalent and therefore we can coalesce them into a single node (the `representative' of the cycle). 


A typical example where cycles occur is when the price of a product is given in two (or more) different currencies, for example in dollars and in pounds, say $\$Price$ and  $\pounds Price$, respectively. In this case, we have the edges $\$Price \to \pounds Price$ and $\pounds Price \to \$Price$ that represent `conversion functions' from the price in dollars to the price in pounds and vice-versa. The existence of these edges makes the two nodes equivalent. Therefore we can choose one of the two nodes as the representative of the equivalence class $\{\pounds Price, \$Price \}$. Incidentally, in this example, one could replace the two nodes by a third, new node $Price$; and if this new node were made clickable then the user could see the `hidden' equivalent nodes by clicking on $Price$. In general, if we click on a node $X$ we see all nodes equivalent to $X$, or the node $X$ alone if there is no node $Y$ different than $X$ but equivalent to $X$. 

Another example where cycles occur is the `attribute renaming' operation of the relational model \cite{Ullman}, where the renaming functions form a cycle containing all attributes that are renamings of a given attribute.

Furthermore, we shall make the following assumptions that we shall need when discussing analytic queries in section \ref{sec:AQ}: 

- every context contains a `terminal node' $T$ such that $dom(T)$ is a singleton

- every node $X$ of a context is equipped with a `terminal edge' $\tau_X: X \to T$  

~~(representing a constant function from $X$ to $T$).




\begin{definition}[Context]
Let $U$ be a universe. A context over $U$ is a finite, labeled, directed, acyclic graph $\mathcal C$ with a single root such that: 
    \begin{itemize}
       \item every node $X$ of $\mathcal C$ is a simple or composite node over $U$, 
       \item every node $X$ of $\mathcal C$ is associated with an edge  $\iota_X: X \to X$ from $X$ to $X$ called the {\em identity edge} of $X$, 
       \item there is a node $T$ with no outgoing edges called the {\em terminal node} of $\mathcal C$,
       \item every node $X$ of $\mathcal C$ is associated with an edge $\tau_X: X \to T$ called the {\em terminal edge} of $X$, 
        \item every simple node of $U$ appears either in the source or in the target of an edge of $\mathcal C$ other than an identity edge (i.e. no isolated nodes).
    \end{itemize}  
\end{definition}

\begin{figure}
{
\begin{center}
\includegraphics[width=350px,keepaspectratio]{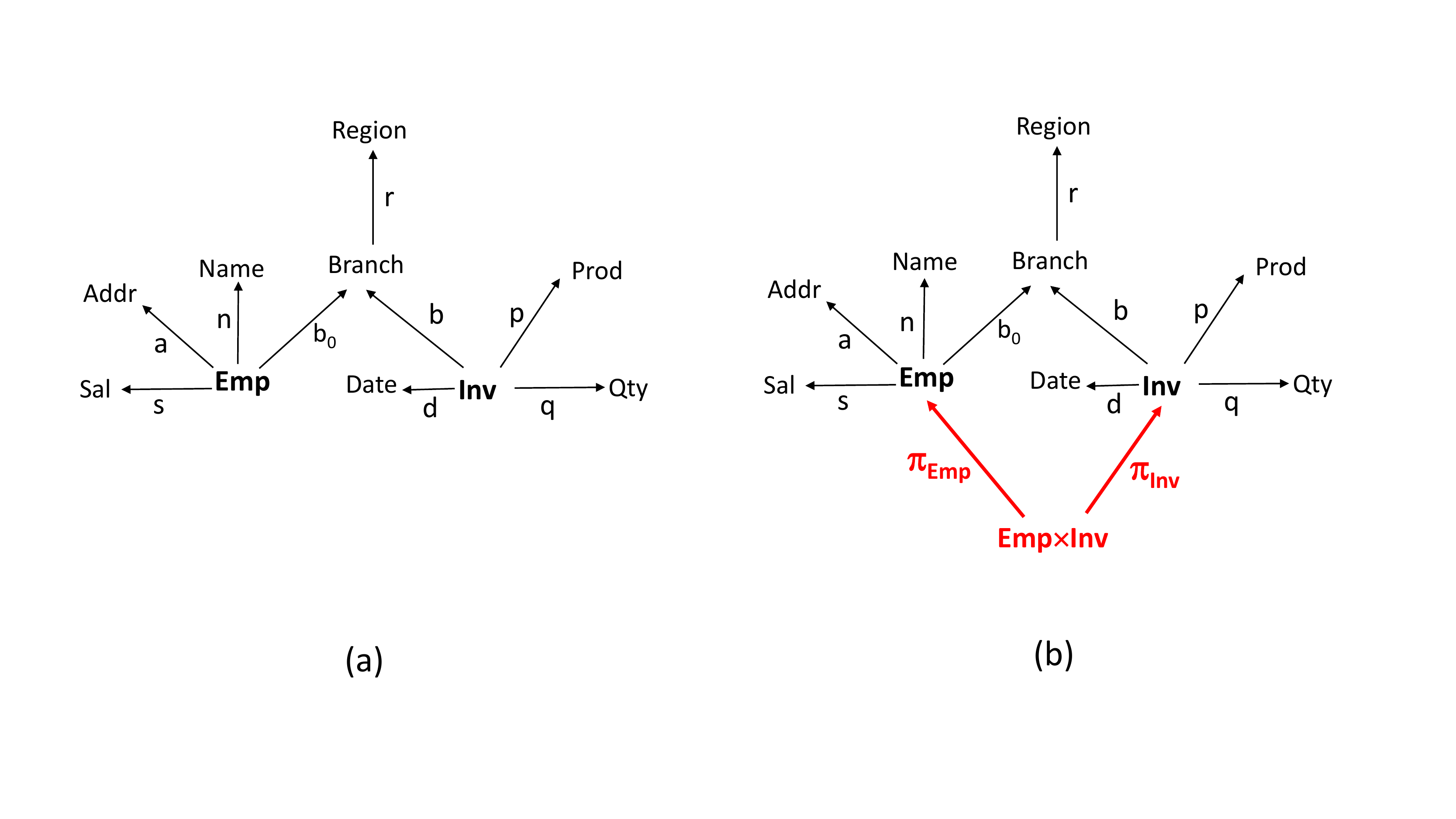}
\caption{(a) Two contexts sharing an edge (b) the two contexts as sub-contexts of a larger context\label{TwoRoots}}
\end{center}
}
\end{figure}

\noindent Some further remarks are in order here regarding the above definition of context: 

\begin{enumerate}
    \item Each node of a context will be assumed equipped with its identity edge, its terminal edge and with its projection edges (if the node is a product node). In our examples, however, we will not show these special edges if not necessary, but we will always assume their existence. \\
    \item Although acyclic, a context is not necessarily a tree. In particular, a context can have parallel paths as in Figure~\ref{CxtDef}(b). Here, the term `parallel paths' is used to mean paths with the same source and same target. More formally, seen as syntactic objects, the edges of a context are triples of the form $\langle source, label, target\rangle$, therefore two edges are different if they differ in at least one component of this triple. This implies, in particular, that two edges can have the same label if they have different sources and/or different targets. Moreover, two different edges can have the same source and the same target provided that they have different labels; we call such edges {\em parallel edges}. Incidentally, it is because of the possibility of having parallel edges that we require that edges be labelled in the above definition (otherwise there would be no way to distinguish between two or more parallel edges). \\
    \item An acyclic graph may have more than one root, so one may wonder why we assume that a context has only one root. The answer is that, as we discussed in the introduction, we want a context to represent the data sets of a {\em single} application and their relationships. This assumption does not prevent the contexts of two or more applications to share nodes and/or edges. Figure \ref{TwoRoots}(a) shows the contexts of two applications, the $Emp$-context and the $Inv$-context, sharing the edge $r$ (i.e. sharing the relationship between branches and regions). Moreover, if necessary, two or more contexts can be `put together' as sub-contexts of a larger context. For example, the contexts of Figure \ref{TwoRoots}(a)  can be put together as sub-contexts of a larger context rooted at $Emp \times Inv$, with the projections $\pi_{Emp}$ and $\pi_{Inv}$ providing the connections of the root  $Emp \times Inv$ to the sub-contexts rooted at $Emp$ and $Inv$, respectively (as shown in Figure \ref{TwoRoots}(b)). Therefore, without loss of generality we can assume that a context has a single root.\\
    \item A context can be seen as the interface between users of an application and the data sets of the application, in the sense that (as we shall see) users formulate their queries using the nodes and edges of the context. Therefore a context plays the role of a schema. However, in contrast to, say, a relational database schema, a context is {\it not} aware of complex structures in data. For example, a context is not aware of how functions might be grouped together to form relations. It is the query language that gives the possibility to users to create such complex structures. We shall come back to this important point in section~\ref{sec:Apps}, where we describe how a relational database can be defined as a view of a context. 
\end{enumerate}

\subsection{The definition of database} 

A context is a syntactic object whose nodes and edges can be associated with data values as discussed in the introduction. These associations constitute what we call a database over a context. 

\begin{definition}[Database]
Let $\mathcal C$ be a context. A database over $\mathcal C$ is a function $\delta$ that associates the nodes and edges of $\mathcal C$ with values such that:
\begin{itemize}
    \item for each simple node $A$ of $\mathcal C$, $\delta(A)$ is a finite subset of $dom(A)$, 
    \item for each product node $ X= A_1 \times \ldots \times A_n$, $\delta(X)= \delta(A_1) \times \ldots \times \delta(A_n)$, $n \geq 2$; and for each
        projection $\pi_{A_i}: X \to A_i$, $\delta(\pi_{A_i})((a_1, \ldots, a_n))= a_i$, for all $(a_1, \ldots, a_n)$ in $X$,
    \item for each edge $f: X \to Y$ of $\mathcal C$, $\delta(f)$ is a  total function from $\delta(X)$ to $\delta(Y)$,
    \item for each node $X$ of $\mathcal C$, $\delta(\iota_X)$ is the identity function on $\delta(X)$; and as a consequence,
        $\delta(\iota_{A \times B})((a, b))= (\iota_A(a), \iota_B(b))$, for all $(a, b)$ in $\delta(A \times B)$; and a similar argument holds for the 
        product of more than two nodes,
    \item $\delta(\tau)$ is a singleton.
\end{itemize}
\end{definition}

It is important to note that a database assigns values {\em only} to simple nodes. Product nodes receive their values from the values of the simple nodes they are composed of. For example, in Figure~\ref{Price}, $\delta(Sup \times Cat)= \delta(Sup) \times \delta(Cat)$.  Indeed, when we form a product of two sets $A$ and $B$ we also define the projection functions $\pi_A: A \times B \to A$ and $\pi_B: A \times B \to B$ which assumes the existence of sets $A$ and $B$ \cite{DBLP:books/daglib/0069193}.  Therefore a product node $A \times B$ cannot be a node in a context unless both, $A$ and $B$, are nodes of the context. In fact here lies the fundamental difference between our model and the relational model. In our model the primitive notion  is that of {\em edge} (i.e. of function) and not that of {\em product node} (i.e. of relation). As we shall see later on, relations in our model are derived notions  defined by queries. \\

\noindent Two remarks are in order here regarding the above definition of database: 

\begin{enumerate}
    \item As the data assigned by $\delta$ to an edge $f: X \to Y$ of a context is a function, and as the current instances of $X$ and $Y$ are finite, the current instance $\delta(f)$ of $f$, is always a finite function. Moreover, as we assume that $\delta(f)$ is a {\em total} function, this means that $\delta(f)$ is defined on every value of $\delta(X)$ which means that we assume {\em no nulls}. 
\item As the terminal node is assigned a singleton set, the function $\delta(\tau_X): \delta(X) \to \delta(T)$ is a constant function, for every node $X$. 

\end{enumerate}





\subsection{The functional algebra}\label{FA} 
In order to access the data stored in a database over of a context users need operations to combine nodes and edges so as to formulate queries. In our model we use one operation to combine sets, namely the {\em Cartesian product}, and three operations to combine functions, namely {\it composition} of two or more functions, {\it restriction} of a function to a subset of its domain of definition, and {\it pairing} of two or more functions with the same source (as defined in the introductory section). These four operations are well known, elementary operations that we call collectively the {\em functional algebra} of a context. 

\begin{definition}[Functional algebra]
Given a context $\mathcal C$, the functional algebra of $\mathcal C$ consists of the following operations: 
\begin{itemize}
    \item {\em Cartesian product} of sets (and its accompanying projection functions).
    \item {\em Restriction}: Given a function $f:X\to Y$ and a subset $S$ of $X$, the restriction of $f$ to $S$, denoted by $f/S$, is the function $f/S: S \to Y$ defined by: $(f/S)(x)= f(x)$, for all $x$ in $S$.
    \item {\em Composition}: Given two functions $f:X \to Y$ and $g:Y \to Z$, their composition, denoted by $g \circ f$ is the function $g \circ f:X \to Z$ defined by: $(g \circ f)(x)= g(f(x))$ for all $x$ in $X$.
    \item {\em Pairing}: Given two functions $f:X \to Y$ and $g:X \to Z$ with common source, their pairing, denoted by $f\wedge g$, is the function $f\wedge g:X \to Y \times Z$ such that: $(f\wedge g)(x)= (f(x), g(x))$ for all $x$ in $X$.
\end{itemize}    
\end{definition} 

\noindent Note that the operations of composition and pairing can be defined for more than two functions in the obvious way. 

\begin{figure}\label{Cxt4}
{
\begin{center}
\includegraphics[width=350px,keepaspectratio]{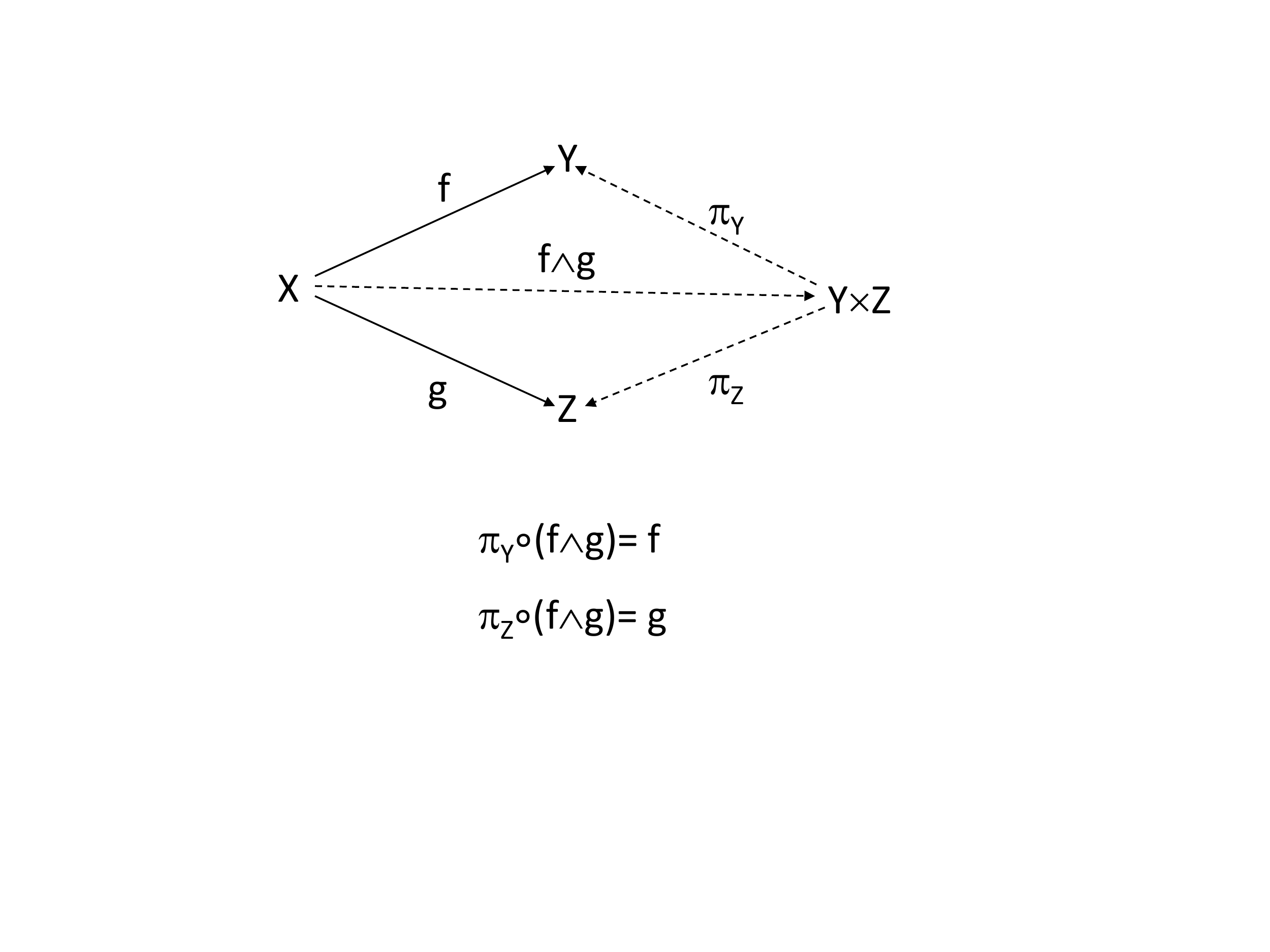}
\caption{Illustrating lemma \ref{BasicLemma}\label{Lemma}}
\end{center}
}
\end{figure}

An important remark concerning the functional algebra is that its four operations are strongly connected to each other as stated in the following lemma. Its proof is a direct consequence from the definitions of the operations (see also Figure~\ref{Lemma}, where $\pi_Y$, $\pi_Z$ denote the projections of $Y \times Z$ over $Y$ and Z, respectively).

\begin{lemma}\label{BasicLemma} Let $X, Y, Z$ be three sets, and let $f: X \to Y$ and $g: X \to Z$ be two functions with common source. Then the following hold: 

\noindent $\pi_Y \circ (f \wedge g)= f$ ~and~ $\pi_Z \circ (f \wedge g)= g$
\end{lemma}

\noindent There is an interesting `derived' operation, called `product' of functions, defined as follows:


\begin{definition}[Product of functions]
Let $f: X \to Y$ and $g: X' \to Y'$ be two functions. The {\em product} of $f$ and $g$ is a function $f \times g: X \times X' \to Y \times Y'$ defined by: $(f \times g)(x, x')= (f(x), g(x'))$ for all $(x, x')$ in $X \times X'$
\end{definition}

\noindent Clearly, the above definition of product can be extended to more than two functions in a straightforward manner. The following lemma states how the product can be derived from operations of the functional algebra, namely using projection, composition and pairing. Its proof follows immediately from the definitions. 

\begin{lemma}
Let $f: X \to Y$ and $g: X' \to Y'$ be two functions. Then we have: $f \times g= (f \circ \pi_X) \wedge (g \circ \pi_{X'})$

\end{lemma}



 \noindent We now introduce the basic concept of `expression' over a context  $\mathcal C$ and its `value' in a database $\delta$ over  $\mathcal C$. 

\begin{definition}[Expression over $\mathcal C$] \label{Expr}
     Let $\mathcal C$ be a context and $\delta$ a database over $\mathcal C$. An {\em expression} over  $\mathcal C$ is either an edge of  $\mathcal C$ or a well formed expression $E$ whose operands are edges of $\mathcal C$ and whose operations are among those of the functional algebra. Moreover, given a database $\delta$ over  $\mathcal C$, the {\em value} of $E$ in $\delta$, denoted as $\delta(E)$, is obtained by (a) replacing the nodes and edges of $E$ with the values assigned to them by $\delta$ and (b) performing the operations.   
\end{definition}

Note that this definition is in the spirit of a similar definition in the relational model, namely the definition of relational expression over a database schema \cite{Ullman}. 

 It is important to note that every expression $E$ over $\mathcal C$ has a source and a target that can be defined recursively based on the sources and targets of the edges appearing in $E$. For example, in Figure~\ref{Price}, if $E_1= r \circ b$ then $source(E_1)= Inv$ and $target(E_1)= Region$; and similarly, if $E_2= (r \circ b) \wedge p$  then $source(E_2)=Inv$ and $target(E_2)= Region \times Product$. Clearly, the value of an expression is always a function from the source of the expression to its target. Moreover, if two expressions have the same source and the same target we shall call them {\em parallel expressions}.

 In the rest of the paper, in order to simplify our discussions, we shall often confuse syntax and semantics that is we shall confuse an expression with its value, when no ambiguity is possible. Here are some examples, referring to Figure \ref{Price}:
 \begin{itemize}
     \item If we say `the function $b$' this will mean the function $\delta(b)$; if we say `the function $r \circ b$' this will mean the function $\delta(r) \circ \delta(b)$; if we say `the function $s \wedge c$' this will mean the function $\delta(s) \wedge \delta(c)$; and if we write $b^{-1}$ this will mean $(\delta(b))^{-1}$. 
     \item If $E$ is an expression and we say `the function $E$' this will mean the value of $E$ (which is a function); and if we write $E^{-1}$ this will mean the inverse of the value of $E$.
     \item If $E$ and $E'$ are two parallel expressions and we write $E= E'$ this will mean that $\delta(E)= \delta(E')$; for example, if we write  $r \circ b= h \circ s \circ p$ this will mean that $\delta(r) \circ \delta(b)= \delta(h) \circ \delta(s) \circ \delta(p)$. 
 \end{itemize}



There is an interesting question concerning the functional algebra. Let $\mathcal C$ be a context over universe $U$, $\delta$ a database over $\mathcal C$ and $f: X \to Y$ a function, where $X, Y$ are (simple or composite) nodes over $U$. We shall say that $\delta$ {\em implies} $f$ if $f$ is the value of an expression over $\mathcal C$. Then the question is: what is the set of all functions that are `implied' by $\delta$? The following lemma gives a first answer. 

\begin{lemma}\label{AA}
 Let $\delta$ be a database over a context $\mathcal C$.  Then the following hold:
 \begin{itemize}
     \item If $X$ is a simple node then  $\delta$ implies the function $\iota_X$; and if $X$ is a product node and $Y$ a sub-product of $X$ then $\delta$ implies the function $\pi_Y: X \to Y$.
     \item If two functions $f:X \to Y$ and $g:Y \to Z$ are implied by $\delta$ then $g\circ f:X \to Z$ is also implied by $\delta$.
     \item If the function $f:X \to Y$ is implied by $\delta$ and $Z$ is a node of $C$ then $f \times \iota_Z:X \times Z \to Y \times Z$ is also implied by $\delta$ (the node Z can be a simple node or a product node).
 \end{itemize}
 \end{lemma}
 
\noindent{\em Proof} The first implication follows from the definition of product; the second follows by applying composition of functions; and the third follows by taking the product of $f$ with the identity function of $Z$. \\ 

\noindent We recall here that the properties expressed in the above lemma correspond to Armstrong's axioms, or implication rules for functional dependencies in relational databases (namely, reflexivity, transitivity and augmentation, respectively)~\cite{DBLP:conf/ifip/Armstrong74}\cite{DBLP:books/cs/Maier83}. However, the above lemma shows only the `soundness' of these implication rules, therefore the accompanying question is the following: is there a `derivation' system showing completeness? Answering this kind of questions lies outside the scope of the present paper. The interested reader is referred to \cite{LS-submitted} for a detailed account of this and other related questions. \\

\noindent We end this section by summarizing a few basic properties of the operations of the functional algebra:

\begin{enumerate}
    \item Strictly speaking, the Cartesian product is neither associative nor commutative. However, there are always natural bijections between the various forms of a Cartesian product, for example, between $A \times B \times C$, $(A \times B) \times C$ and $A \times (B \times C)$, or between $A \times B$ and $B \times A$, and so on. Therefore, in our setting, we shall assume that the Cartesian product is both associative and commutative. As a consequence of this assumption, pairing is also associative and commutative. 
    \item Composition is an associative operation, therefore we can group together its component functions as it is more convenient for our purposes. For example, in Figure \ref{Price}, $h \circ s \circ p= (h \circ s) \circ p= h \circ (s \circ p)$
    \item The restriction of the pairing of two functions $f:X \to Y$ and $g:X \to Z$ on a subset $S$ of $X$ equals the pairing of the restrictions of the two functions: $(f \wedge g)/S= (f/S) \wedge (g/S)$
    \item The composition of restricted functions has the following properties, for any functions $f:X \to Y$ and $g:Y \to Z$: 
    
    - if $S \subseteq X$, then $g \circ (f/S)= (g \circ f)/S$ 
    
    - if $S \subseteq X$ and $T \subseteq Y$ then $(g/T) \circ (f/S)= (g \circ f)/W$, where $W=S \cap f^{-1}(T)$ 
    
    ~~(in other words, in this case, the restriction $T$ of $Z$ is `pushed' backwards as

    ~~a restriction $W$ of $X$).
    \item In view of the previous item, consider the composition of $n$ restricted functions forming a path as follows, where 
    $S \subseteq X$ and $S_i \subseteq X_i$, $i= 1, \ldots, n$:
    $$X/S \xrightarrow{f_1} X_1/S_1 \xrightarrow{f_2} X_2/S_2 \ldots X_{n-1}/S_{n-1} \xrightarrow{f_n} X_n/S_n$$ 
    Then the restrictions $S_i$ can be `pushed' to the source of the path as a restriction $W_1$ of $X$ computed as follows: 

    $W_n= S_{n-1} \cap f_n^{-1}(S_n)$, ~$W_{n-1}= S_{n-2} \cap f_{n-1}^{-1}(W_n)$, \ldots, $W_1= S \cap f_1^{-1}(W_2)$\

\end{enumerate}


\subsection{Integrity constraints}\label{IC}

Integrity constraints are properties that the database must satisfy so that the quality of information is maintained. Examples of such constraints from relational databases are domain constraints, key constraints, or referential constraints \cite{Ullman}. Integrity constraints ensure that changes in the database are performed in such a way that data integrity is not affected. 

In our model we distinguish two kinds of constraints, namely those imposed by the definition of database over a context, and those that may be imposed by the application. There are two constraints imposed by our definition of database, namely: (a) the database must associate each node $X$ with a {\em finite} subset of $dom(X)$ and (b) the database must associate each edge $f: X \to Y$ with a {\em total} function  $\delta(f): \delta(X) \to \delta(Y)$ (meaning that no nulls are allowed in the database). 

\noindent Incidentally, the second constraint implies a referential constraint similar to referential constraints in the relational model. Indeed as each edge $e: X \to Y$ of a context is interpreted as a total function,  every object of $\delta(X)$ must `refer' to an object of $\delta(Y)$ (i.e. no `dangling pointers').

As for constraints that may be imposed on a database by the application, we consider two kinds, namely {\em equality constraints} and {\em refinement constraints} (also called {\em dependency constraints}).  An equality constraint requires that two or more parallel expressions have the {\em same} value in the database, whereas a  refinement constraint requires that an expression $E$ contain {\em finer information} than an expression $E'$ with the same source as $E$. 

\begin{definition}[Equality constraint]
    Let   $\mathcal C$ be a context, $\delta$ a database over  $\mathcal C$ and $E, E'$ two parallel expressions over  $\mathcal C$ . We say that $\delta$ {\em satisfies} the equality constraint $E= E'$ if $\delta(E)= \delta(E')$. 
\end{definition}

\noindent As a shorthand, we shall often say that `$\delta$ satisfies $E= E'$' instead of saying `$\delta$ satisfies the equality constraint $E= E'$'; and similarly, we shall say that `$\delta$ satisfies equalities' instead of saying `$\delta$ satisfies all equality constraints'.


Note that the concept of equality constraint of our model is strongly related to that of functional dependency in the relational model. To see this consider a relation schema $R(A, B, C)$ with functional dependencies $f:A \to B$, $g:B \to C$ and $h:A \to C$. Then it is easy to see that a relation $r$ over $R$ satisfies these three dependencies if and only if  $g \circ f= h$. Indeed  if r satisfies $f$ and $g$, then $r$ must also satisfy  the function $g \circ f= A \to C$  (by the transitivity property of functional dependencies in the relational model). It follows that $r$ must satisfy two functions from $A$ to $C$, namely $h$ and $h'$. Therefore  $r$ will be consistent only if $h= h'$ otherwise the projection of $r$ over $AC$ may not be a function. 

 The second constraint that we consider in our model is what we call a {\em refinement constraint} or {\em dependency constraint}. It is an information-theoretic constraint referring to the information content of expressions having the same source. To understand the definition of this constraint, recall that an expression $E$ over a context represents a total function and therefore its information content can be defined as the partition $p_E$ that the function $E$ induces on its domain of definition. It follows that we can compare two expressions having the same source using the ordering of the partitions they induce on their common source. 

\begin{definition}[Refinement constraint]\label{RC}
Let $\mathcal C$ be a context and let $E, E'$ be expressions over $\mathcal C$ having the same source. Then $E$ is said to be {\em finer} than or equal to $E'$, denoted by $E \leq E'$, if $p_E \leq p_{E'}$. Moreover, if $E \leq E'$, we shall often say that $E$ {\em determines} $E'$, denoted by $E \to E'$.  
\end{definition}

Note that an equality constraint is equivalent to two  refinement constraints, since for any two expressions $E, E'$ we have that: $E= E'$ if and only if $E \leq E'$ and $E' \leq E$. This means that we could consider just one constraint in our model, namely the refinement constraint. However, as we shall see, it is more `convenient' for our discussions to keep both. The following lemma states an interesting property relating the two constraints. 

\begin{lemma}\label{EdgeDep}
 Let $\mathcal C$ be a context and let  $f: X \to Y$ and $g: X \to Z$  be two edges of $\mathcal C$ having the same source $X$.   Then $f \leq g$ if and only if there is a function $h: Y \to Z$ such that $h \circ f = g$. Moreover, the function $h$ is unique.
\end{lemma}
\proof  Suppose first that $f$ determines $g$, that is $p_f \leq p_g$, and define $h(y)= g(f^{-1}(y))$ for every $y$ in the range of $f$ (i.e. in the set of values of $f$). Then $h$ is a well defined function from $Y$ to $Z$. Indeed, as $p_f \leq p_g$, every block of $p_f$ is contained in a block of $p_g$ and therefore $g(f^{-1}(y))$ is a single value in the range of $g$. It follows that $h:range(f) \to Z$ is a well defined function and that $h \circ f = g$. 

\noindent In the opposite direction, suppose that $h \circ f = g$. It follows that $p_{h \circ f} = p_g$ and as $p_f \leq p_{h \circ f}$ we have that $p_f \leq p_g$. 

\noindent Now, suppose that there is a function $h':Y \to Z$ such that $h \circ f = g'$. If $h' \ne h$ then there is $y \in range(f)$ such that $h'(y) \ne h(y)$. It follows that there is  $x \in X$ such that $f(x)= y$, therefore $h'(f(x)) \ne h(f(x))$, that is $g(x) \ne g(x)$ which is impossible as $g$ is a function. \\ 

\noindent Clearly the proof remains the same if $f, g$ and $h$ in the above lemma are expressions over $\mathcal C$ (since the value of an expression in a database over $\mathcal C$ is a function).

\medskip
To see an example of refinement constraint refer to Figure~\ref{CxtDef}(a) and consider the edges $c$ and $s$ that associate each product with a category and a supplier. If we declare the refinement constraint $c \leq s$ this means that each product category must be provided by one and only one supplier. In other words, a refinement constraint works in much the same way as a functional dependency in the relational model. Let's see a simple example illustrating this point. Consider the following context relating employees, departments and managers in some enterprise: 

$\mathcal C= \{f:Emp \to Dep, g:Dep \to Mgr, h:Emp \to Mgr\}$

\noindent The path $Emp \to Dep \to Mgr$ means that the manager of an employee is the manager of the employee's department, whereas the path $Emp \to Mgr$ means that the manager of an employee is assigned directly to the employee without reference to the employee's department. Imposing the equality constraint $g \circ f= h$ is tantamount to saying that the manager of an employee must be the same as the manager of the employee's department. Note that if we consider the context  $\mathcal C'= \{f:Emp \to Dep, h:Emp \to Mgr\}$ we could have the same result by imposing the refinement constraint $f \leq h$.

We shall say more on equality constraints and refinement constraints in section~\ref{Embedding}. Our intention in this section was to give a brief account of the kind of constraints that we use in our model. A detailed discussion of constraints and their inference mechanism lies outside the scope of the present paper and it is part of our current work. \\

 \noindent We have seen so far the basic concepts of the graph database model that we propose in this paper, namely the concepts of context and database over a context; the functional algebra for manipulating the edges of a context; and the basic constraints that can be imposed on a database. In the following two sections we define the query language of our model. It consists of two distinct types of queries, namely {\em traversal queries} and {\em analytic queries}. As we have already mentioned, a distinctive feature of our model is that both types of queries are defined within the same formal framework (i.e. within the functional algebra) - in contrast to the relational model where analytic queries are defined outside the relational algebra.

\section{Traversal Queries}\label{sec:TQ}

In this section we define traversal queries and their evaluation techniques - in particular rewriting techniques, and we discuss the concept of view in our model. 
 

\subsection{The definition of a traversal query}\label{TQDef}

A traversal query is a type of query used to explore and navigate a graph data structure by traversing its nodes and edges, starting from a specific node, or set of nodes, and following the connections between nodes based on defined criteria. \\

\noindent Our definition of traversal query relies on the concept of `expression' over a context $\mathcal C$ introduced earlier (see section  \ref{FA}). Clearly, any such expression can be seen as a query over  $\mathcal C$ and its value in a given database $\delta$ as its answer in $\delta$. However, in our approach, we choose to call traversal query a specific type of expression, namely the pairing of any number of expressions over  $\mathcal C$ having common source. The reason for this choice is that if we want to explore a graph starting at some (simple or composite) node, then we have two options: either follow a path or explore two or more paths in parallel and pair their values.

\begin{definition}[Traversal query] Let $\mathcal C$ be a context. A {\em traversal query} $Q$ over $\mathcal C$ is either a single expression $E$ over $\mathcal C$ or the pairing $E_1 \wedge \ldots \wedge E_n$ of two or more expressions over $\mathcal C$, having the same source, say $K$.  Moreover, if $\delta$ is a database over $\mathcal C$ then the {\em answer} of $Q$ in $\delta$, denoted by $Ans(Q, \delta)$, is the value of the expression $Q$ in $\delta$; that is, the answer of $Q$ in $\delta$ is a function defined as follows:

$Ans(Q, \delta): \delta(K) \to \delta(E_1) \times \ldots \times \delta(E_n)$ such that:  

$(Ans(Q, \delta))(k)= (\delta(E_1) \wedge \ldots \wedge \delta(E_n))(k)$, for all $k$ in $\delta(K)$ 
\end{definition} 
\smallskip
\noindent We shall call the common source $K$ of the expressions in a traversal query the {\em key} of the query, and we shall use the notation $Q(K, E_1, \ldots, E_n)$ whenever we want to show explicitly the key of the query.  Therefore, we will use two equivalent notations to denote a traversal query: $Q= E_1 \wedge \ldots \wedge E_n$ and $Q(K, E_1, \ldots, E_n)$. Moreover, we shall often use the term `query' instead of  `traversal query', and we shall also denote the answer of a query $Q$ simply by $Ans_Q$ when $\delta$ is understood. \\

The simplest forms of traversal query are: a single edge, a single path or the pairing of two paths. For example, $Q=r$, $Q'= r \circ b$ and $Q''= (r \circ b) \wedge (c \circ p)$ are traversal queries over the context of  Figure \ref{Price}.  \\ 

Note that the key of a query can be a simple node or a composite node. For example, in Figure \ref{Price}, the query $Q= r \circ b$ has the simple node $Inv$ as key, whereas the query $Q= u$ has the composite node  $Sup \times Cat$ as key. As for the target of a query, it can be computed recursively as discussed earlier (see section \ref{FA}). For example,  $Q= (r \circ b) \wedge (s \circ p)$ is a traversal query, whose expressions are $r \circ b$ and $s \circ p$. Its key is $Inv$ and its target is  $Branch \times Sup$. Note that, following our convention, we can also denote this query as  $Q(Inv, r \circ b, s \circ p)$. \\

\noindent In the rest of this section we discuss several important facts about traversal queries that we shall use later. We illustrate them through examples, using the context of Figure \ref{Price}. 
 
\begin{enumerate}
    \item The {\em only} restriction imposed on the expressions that we can use in a query is that they all must have  common source, which is the key $K$ of the query. However, two or more expressions of a query can be parallel (i.e. they can have the same target). For example, the two expressions of the query $Q(Inv, r \circ b, h \circ s \circ p)$ are parallel, as both have $Region$ as their target. \\
    \item The target of an expression in a query can be a product node. For example, the target of the second expression of the query $Q(Inv, b, (s \wedge c) \circ p)$  is $Sup \times Cat$, which is a product node. However, for simplicity, and without loss of generality, we will assume that the targets of all expressions in a query are simple nodes. Indeed, if the target of an expression is a product node then we can replace the expression by as many expressions as there are factors in the product node - each expression having a simple node as target. This is possible using the following lemma, whose proof is an immediate consequence of the definitions of operations in the functional algebra.


\begin{lemma}\label{Grouping} Let $X, Y, Z$ be three nodes of a context $\mathcal C$ and $f: X \to Y$, $g: Y \to Z$ and $h: Y \to W$ three edges of $\mathcal C$. Then we have: $(g \wedge h) \circ f= (g \circ f) \wedge (h \circ f)$
\end{lemma}

Following the above lemma, we can replace the single expression of the query $Q(Inv, (s \wedge c) \circ p)$ by the pairing of the expressions $s \circ p$ and $c \circ p$ to obtain the `equivalent' query $Q'(Inv, s \circ p, c \circ p)$, in which each expression has a simple node as target. Here, query equivalence is defined as follows: two queries  $Q$ and $Q'$ over a context $\mathcal C$ are equivalent, denoted as $Q \equiv Q'$, if they return the same answer in every database $\delta$ over $\mathcal C$. 

\noindent Note that the possibility to define a traversal query using only expressions whose targets are simple nodes leads to a convenient way for defining a traversal query: (a) choose a node $K$ of $\mathcal C$ as the key, (b) choose $n$ simple nodes $A_1, \ldots, A_n$ of $\mathcal C$ as expression targets and (c) for each $i$ give one or more expressions with $K$ as the source and $A_i$ as the target. We shall come back to this point when discussing the use of a context as an interface (section \ref{subsec:Interface}). \\
\item As a traversal query $Q(K, E_1, \ldots, E_n)$ is a pairing, its answer, $Ans_Q$, can be seen as a set of tuples over the attributes  $K, A_1, \ldots, A_n$, where $A_1= target(E_1), \ldots, A_n= target(E_n)$. For example, consider  the query $Q(Inv, r \circ b, s \circ p)$, in which the expression $r \circ b$ has $Region$ as target, and  the expression $s \circ p$ has $Sup$ as target. The answer to this query can be seen as a set of tuples over the attributes $Inv, Region$ and  $Sup$. In other words, each traversal query $Q(K, E_1, \ldots, E_n)$ {\em induces} a relation schema $R_Q(K, A_1, \ldots, A_n)$, and for every database $\delta$, the answer $Ans(Q,\delta)$ {\em induces} a relation $r_Q$ over $R_Q$. Moreover, as $\pi_{A_i} \circ (Ans(Q, \delta))$ is a function from $K$ to $A_i$, $i= 1, \ldots, n$, the relation $r_Q$ {\em satisfies} the functional dependencies $K \to A_1, \ldots, K \to A_n$. It follows that there is a mapping from traversal queries $Q$ over $\mathcal C$ to relation schemas $R_Q$ such that: (a) for each database $\delta$ over  $\mathcal C$, the answer $Ans(Q, \delta)$ induces a relation $r_Q$ over $R_Q$ satisfying the functional dependencies $K \to A_1, \ldots, K \to A_n$, and (b) the values of each non-key attribute $A_i$ of $R_Q$ are computed by the expression $E_i$ having $A_i$ as its target. In other words, the expression $E_i$ of $Q$ provides the {\em semantics} of attribute $A_i$ of $R_Q$, $i= 1, \ldots, n$.

\noindent However, an issue arises here if two expressions, $E$ and $E'$, are parallel that is if they have the same target, say $A$. In this case there is ambiguity as the two expressions compute in general different values in the domain of $A$. One solution is to `split' the common target $A$ into two targets, denoted as $E.A$ and $E'.A$, such that the prefix indicates which expression is used for computing which values of the domain of $A$. For example, consider the query $Q(Inv, r \circ b, h \circ s \circ p)$, whose expressions are $E= r \circ b$ and $E'= h \circ s \circ p$, both having $Region$ as target. In this case, the induced relation schema will be $R_Q(Inv, E.Region, E'.Region)$, with $dom(E.Region)= dom(E'Region)= dom(Region)$, and with clear semantics for the attributes $E.Region$ and $E'.Region$. We shall come back to this point when discussing the definition of a relational database as a view of a context in section  \ref{Embedding}.   \\
\item A particularly interesting kind of traversal query is one in which there are no parallel expressions, and moreover each expression has a simple node as target. In this case the query is a tree, therefore we shall refer to it as a {\em tree query}. A main feature of a tree query is that its answer can be represented in an easy-to-read table. Indeed, if $A_1, \ldots, A_n$ are the targets of the expressions of a tree query, then the answer can be represented in a table whose columns are $K, A_1, \ldots, A_n$ and in which, for each $k$ in $K$, the value $E_i(k)$ is entered in the cell $(k,A_i)$. We recall that, in the relational model \cite{Ullman}, a table where the entries are atomic values and in which the values in each column are functionally dependent on the key-values is said to be in First-Normal-Form. Moreover, if we view each expression $E_i$ of a tree query as a functional dependency (from the key of the query to the target of $E_i$) then the table just defined satisfies all key dependencies. Therefore, as seen in the previous item, the answer to a tree query can be seen as a relation with key $K$ (the key of the query) and with attributes the targets  $A_1, \ldots, A_n$ of the expressions $E_1, \ldots, E_n$ of the query. By the way, any traversal query can become a tree query if we impose the equality constraint (i.e. if we require that any two parallel expressions evaluate to the same function).  We shall come back to these important points in section \ref{Embedding} when defining a relational database as a view over a context. 

\noindent It is worth noting here that a node $X$ (whether simple or composite), together with its projection functions defines a particular type of tree query that we call {\em identity query}. For example, the node $Sup \times Cat$ in Figure~\ref{Price} together with its projection functions, $\pi_{Sup}$ and $\pi_{Cat}$, defines an identity query over the node $Sup \times Cat$. Its answer is the identity function  $\iota_ {Sup \times Cat}$. Typically, identity queries are used when we want to `read' the content of a node $X$. For example, the identity query $Q=\iota_{Sup}$ will return the set $\{(s, s) / s \in \delta(Sup)\}$.   \\
\item As a query is an expression over  $\mathcal C$, we can use the operations of the functional algebra to define new queries from old (in other words there is a notion of `closure' of the set of traversal queries). In particular:

{\em Restriction}: If $Q$ is a query with key $K$ and $S \subseteq K$, then  $Q/S$ is a query with key $S$ and the same target as $Q$. For example if $Q= b \wedge (s \circ p)$ and $S \subseteq Inv$ then  $Q/S$ is a query.

{\em Composition}: If $Q'$ and $Q''$ are queries such that target$(Q')$= source$(Q'')$, then the composition $Q'' \circ Q'$ is a query whose  key is the key $Q'$ and whose target is the target of $Q''$. For example if $Q'= s \circ p$ and $Q''= h$ then $Q'' \circ Q'= h \circ (s \circ p)$  is a query with $Inv$ as its key and $Region$ as its target. Regarding the use of the special edges $\iota_A$ and $\tau_A$ in the composition of traversal queries we note the following: for every query $Q$ with source $X$ and target $Y$ we have: ~$\iota_Y \circ Q= Q\circ \iota_X = Q$  ~and ~~$\tau_Y \circ Q= \tau_X$

{\em Pairing} The pairing of two or more queries with common key is a query.  For example, consider the queries $Q'(Inv, r \circ b$)  and $Q''(Inv, p)$. Then their pairing $Q' \wedge Q''$ is a query $Q(Inv, r \circ b, p)$ with $Inv$ as its key and $Region \times Prod$ as its target. Note that the queries in a pairing may have one or more expressions in common. For example, consider the queries $Q'= b \wedge (s \circ p)$ and $Q''= (s \circ p) \wedge (c \circ p)$, which share the expression $s \circ p$, and let  $Q= Q' \wedge Q''$ be their pairing. Then the key of $Q$  is $Inv$ (which is the common key of $Q'$ and $Q''$) and the target of $Q$ is $(Branch \times Sup) \times (Sup \times Cat)$ (which is the product of the targets of $Q'$ and $Q''$). Note that, in this case, the induced relation $R_Q$ will have four attributes, $Branch, Sup, Sup, Cat$, and in order to distinguish between the repeated attribute $Sup$ we will have to use the query name as a prefix. In other words, the relation schema induced by the query $Q= Q' \wedge Q''$ is $R_Q(Branch, Q.Sup, Q'.Sup, Cat)$. Note that this kind of prefixing is the usual practice when taking joins in the relational model \cite{Ullman}. \\
\item It is important to note that equality constraints and refinement constraints can be used either as constraints that every database over a context must satisfy or as constraints that the answer to a query must satisfy. 
For example, consider the following queries:

\noindent $Q1= \pi_{Sup, Region}(r \circ b) \wedge (h \circ s \circ p)$

\noindent $Q2= \pi_{Sup, Region}(((r \circ b) \wedge (h \circ s \circ p))/S)$ 

\hspace{1cm}where $S= \{i \in Inv / (r \circ b)(i)= (h \circ s \circ p)(i)\}$ 

\noindent If the database does not have to satisfy equality constraints then $Q1$ returns a set of pairs $(Sup, Region)$ which may contain pairs $(Sup, Region)$ such that the supplier's region is not the same as the branch's region; whereas $Q2$ returns a pair $(Sup, Region)$ {\em only if} the supplier's region is equal (i.e. the same) as the branch's region. If the database satisfies the equality constraint then $Q1$ and $Q2$ both return the same answer. 
\end{enumerate}

\subsection{Traversal query evaluation}\label{TQEval}

The evaluation of a traversal query can be done directly, based on the definitions of the operations of the functional algebra, or indirectly, either by rewriting the query or by translating it as a query in some query engine for its evaluation (e.g. by translating it as an SQL query). In this section we discuss briefly the evaluation of a traversal query by rewriting. 

The kind of rewriting that we consider is the process of applying a number of transformations to the original query in order to produce an equivalent optimized one (`optimized' in the sense that the equivalent query allows to reuse previously obtained query results). Such transformations do not depend on the physical state of the system (such as the size of physical structures, the system workload, etc). They are just well-defined rules that specify how to transform a query expression into a logically equivalent one \cite{DBLP:reference/db/Pitoura18d}. 

The main rewriting rules that we use are based on properties of the functional algebra. We call them 
{\em associative rule}, {\em distributive rule}, {\em grouping rule}, and {\em restriction propagation rule}.  \\

\textbf{Associative rule }:  $Q_1 \circ Q_2 \circ Q_3 \equiv Q_1 \circ (Q_2 \circ Q_3) \equiv (Q_1 \circ Q_2) \circ Q_3$, where:

\hspace{3cm} $target(Q_1)= source(Q_2)$  and  $target(Q_2)= source(Q_3)$

\smallskip\noindent This rule is based on the associativity of composition in the functional algebra. It means that if we have a query $Q$ which is the composition of more than two sub-queries then we can parenthesize in any way which is more convenient for the evaluation of $Q$. For example, suppose we want to evaluate the query $Q= h \circ s \circ p$, over the context of Figure~\ref{CxtDef}(b). This query is the composition of three sub-queries, $h$, $s$ and $p$, and we can parenthesize it in two different ways, as follows: 

$Q= h \circ s \circ p \equiv h \circ (s \circ p)\equiv (h \circ s) \circ p$ 

\noindent If we have already evaluated the sub-query $s \circ p$ (and have stored the result in a cache) then the second form above is preferable for the evaluation of $Q$; and similarly, if instead of $s \circ p$ we had already evaluated  the subquery $h \circ s$ then the third form would be preferable. 

\noindent The following two rules use the property of the functional algebra expressed by Lemma~\ref{Grouping}. \\

\textbf{Distributive rule}:  $Q \circ (Q' \wedge Q'') \equiv (Q \circ Q') \wedge (Q \circ Q'')$, where: 

\hspace{3cm} $source (Q')= source(Q'')$ and $target(Q' \wedge Q'')= source(Q)$

\smallskip\noindent For example, referring to Figure 1(b), consider the query $Q= p \circ (s \wedge c)$. Then applying the distributive rule we can rewrite $Q$ as $Q'= (s \circ p) \wedge (c \circ p)$. 

\noindent Clearly, this rule can also be used in the opposite direction giving rise to the following rule: \\

\textbf{Grouping rule}: $(Q \circ Q') \wedge (Q \circ Q'') \equiv Q \circ (Q' \wedge Q'')$, where:

\hspace{3cm} $source (Q')= source(Q'')$ and $target(Q' \wedge Q'')= source(Q)$
 
\smallskip\noindent For example, consider the query $Q= (s \circ p) \wedge (c \circ p)$ over the context of Figure~\ref{CxtDef}(b). Applying the grouping rule we can rewrite the query as  an equivalent query $Q'= (s \wedge c) \circ p$, thus `factoring out' $p$ and grouping $s$ and $c$.

\smallskip
The above rules concern queries using composition and pairing. Regarding the use of restriction in queries, we have a rule that `pushes' all restrictions in the query towards the key of the query. We call this rule the `restriction propagation rule': \\

\textbf{Restriction propagation rule}: In order to define this rule, we consider two cases: 
\begin{enumerate}
    \item The query is a single expression $E$ in which some or all of its nodes are restricted. In this case we push all restrictions to the source of $E$ as described by the property of the functional algebra expressed by item 5 of section \ref{FA}.
    \item The query is the pairing of two or more expressions. In this case we push all restrictions in each expression to the source of the query and apply the property of the functional algebra expressed by item 3 of section \ref{FA}.
\end{enumerate}
%
 We note that restricting the nodes of a query corresponds to the `selection' operation in relational algebra queries (hence to the `where' clause of SQL).\\
 
\noindent As a final remark on rewriting,  the general approach is: if you have a query and you want to put it in a specific form (e.g. in order to reuse query results from a cache) then try to do so using properties of the operations of the functional algebra. Clearly, having discovered a set of rewriting rules such as the above helps doing so faster.

\subsection{Views}\label{Views} 







\begin{figure}
{
\begin{center}
\includegraphics[width=350px,keepaspectratio]{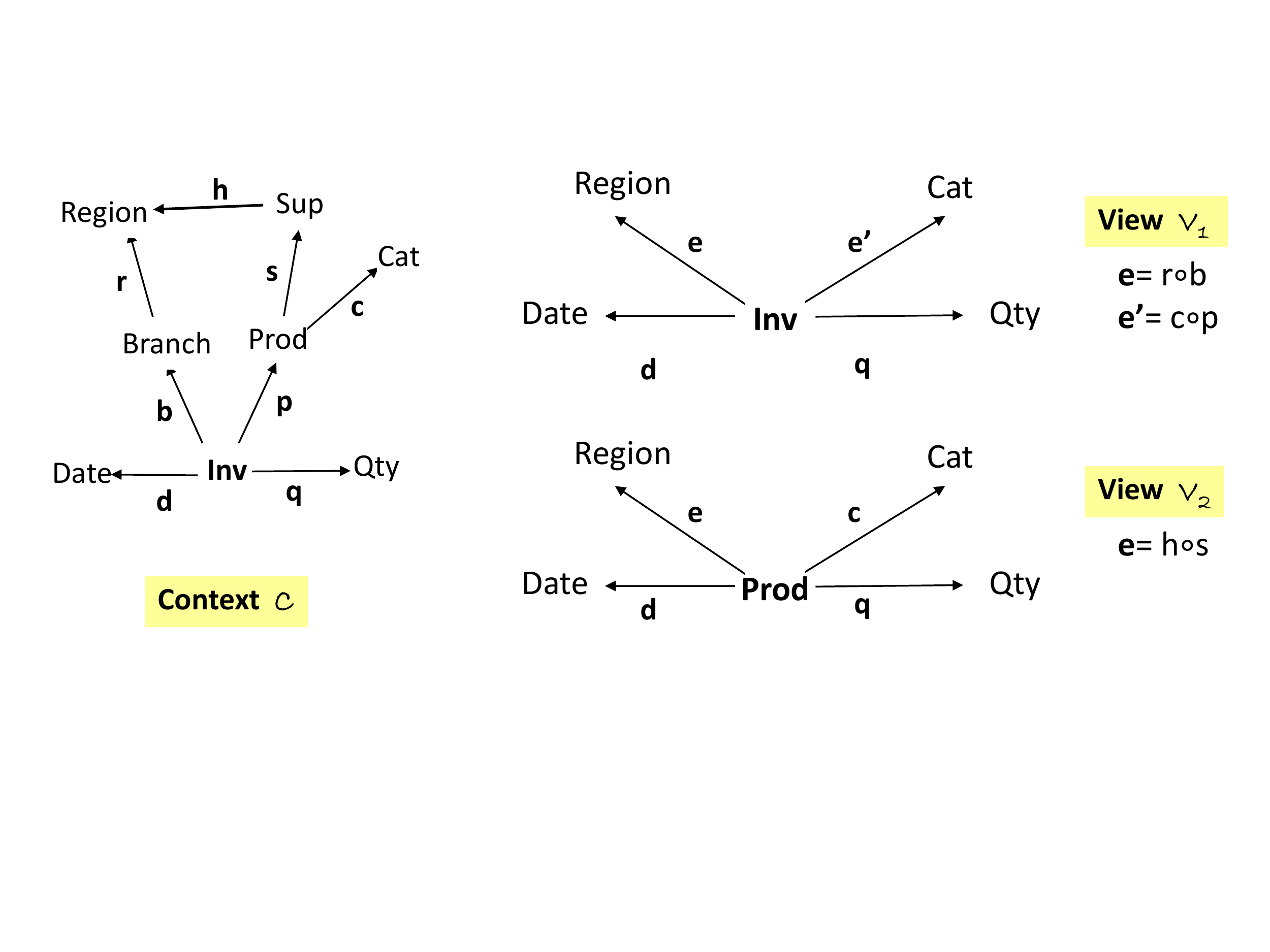}
\caption{A context $\mathcal C$ and two views of $\mathcal C$ with different roots}\label{Fig-Views}
\end{center}
}
\end{figure}

A user or a group of users may want to use only part of the information contained in a context, and may also want that part to be structured again as a context reflecting specific user needs. For example, Figure~\ref{Fig-Views} shows a context $\mathcal C$ and two views of $\mathcal C$, named $\mathcal V_1$ and $\mathcal V_2$. Each edge of $\mathcal V_1$  is either an edge of $\mathcal C$ (such as $d$ or $q$) or a query over $\mathcal C$ (such as the edges $e$ and $e'$). In other words, $\mathcal V_1$ is a context whose edges are queries over $\mathcal C$ (and similarly for $\mathcal V_2$). 

\begin{definition}[View of a context]
 Let $\mathcal C$ be a context. A {\em view} of $\mathcal C$ is defined to be a context $\mathcal V$ whose edges are queries over $\mathcal C$. Given a database $\delta$ over $\mathcal C$, the current instance of $\mathcal V$ is defined to be the set of answers to the queries defining the edges of $\mathcal C$.
 \end{definition}

 It should be clear that the concept of view as well as the problems related to view management in our model are similar to those in the relational model: a view can be virtual or materialized; a query over a virtual view $\mathcal V$ must be translated as a query over the context $\mathcal C$ in order to be answered, whereas a query over a materialized view can be answered directly from the view; when the database over the context  $\mathcal C$ is updated, the updates are propagated to the view only if the view is materialized; updating through views is problematic; and so on (see \cite{Ullman}).  

 As an example consider the view $\mathcal V_1$ in Figure~\ref{Fig-Views}  and the following query over that view: $Q= e \wedge e'$. To answer $Q$, $e$ and $e'$ are replaced by their definitions  to obtain the query $\overline Q= (r \circ b) \wedge (c \circ p)$ which is evaluated over $\mathcal C$ to obtain the answer to $Q$. Note that, as this example shows, the edges of a view of a context $\mathcal C$ can be seen as macros that facilitate the formulation of queries over $\mathcal C$.

\section{Analytic queries}\label{sec:AQ}

In the previous section we defined and discussed the class of traversal queries. In this section we define and discuss the class of analytic queries. These two types of queries make up the query language of the context model.  

\subsection{The definition of analytic query}\label{DAQ}

In order to give a general definition of analytic query over a context we build upon the approach of \cite{SpyratosS18}. 
First, let us recall the definition of analytic query of \cite{SpyratosS18} using our running example. 
Suppose that we want to know the total quantity delivered to each branch during the year. This computation needs only two among the functions shown in Figure \ref{CxtDef}(b), namely $b$ and $q$. Figure 6 shows a toy example of the data returned by $b$ and $q$, where the data set $Inv$ consists of seven invoices, numbered 1 to 7. In order to find the total quantity by branch we proceed in three steps as follows: \\

\noindent{\em Grouping}: During this step we group together all invoices referring to the
same branch (using the function $b$). We obtain the following groups of invoices (also shown in the figure):

– Branch-1: 1, 2

– Branch-2: 3, 4

– Branch-3: 5, 6, 7\\

\noindent {\em Measuring}: In each group of the previous step, we find the quantity corresponding to each invoice in the group (using the function $q$):

– Branch-1: 200, 100

– Branch-2: 200, 400

– Branch-3: 100, 400, 100\\

\noindent {\em Aggregation}: In each group of the previous step, we sum up the quantities found:

– Branch-1: 200 + 100 = 300

– Branch-2: 200 + 400 = 600

– Branch-3: 100 + 400 + 100 = 600 \\

\begin{figure}
{
\begin{center}
\includegraphics[width=350px,keepaspectratio]{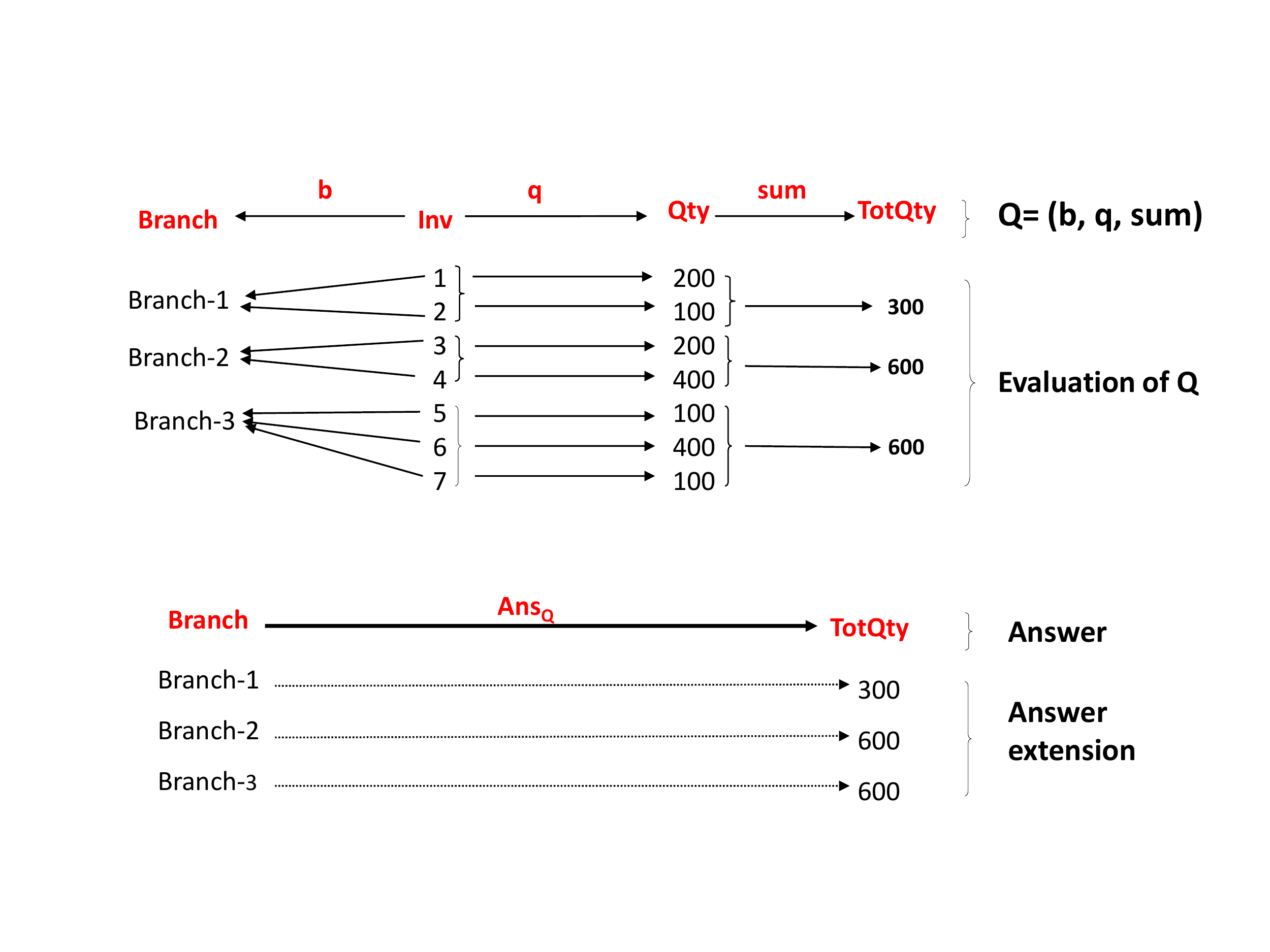}
\caption{The evaluation of an analytic query\label{Eval}}
\end{center}
}
\end{figure}

\noindent Then the association of each branch to the corresponding total quantity, as shown in Figure \ref{Eval}, is the desired result:

– Branch-1 → 300

– Branch-2 → 600

– Branch-3 → 600\\

We view the ordered triple $Q= (b, q, sum)$  in Figure \ref{Eval} as an analytic query over the context, the function $Ans_Q: Branch \rightarrow TotQty$ in that same figure as the answer to $Q$, and the computations shown in the figure as the query evaluation process.
Note though that what makes the association of branches to total quantities possible is the fact that: (a) the functions $b$ and $q$ have a common source (namely, $Inv$ in this example), and (b) that `sum' is an operation applicable on $q$-values. Also note that {\em TotQty} is a new symbol necessary for naming the co-domain of the computed function $Ans_Q$.

The function $b$ that appears first in the triple $(b, q, sum)$ and is used in the grouping step is called the {\it grouping function}; the function $q$ that appears second in the triple is called the {\it measuring function}, or the measure; and the function $sum$ that appears third in the triple is called the {\it aggregate operation}. Actually, the triple $(b, q, sum)$ should be regarded as the specification of an analysis task to be carried out over the data set $Inv$. 

Note that exchanging the two first components of this triple we obtain the query $(q, b, sum)$ which is {\em not} a well formed query as the aggregate operation is not applicable on b-values which are Branches (i.e. we can't sum up branches). However if instead of `sum' we put `count' as the aggregate operation then we obtain the query $(q, b, count)$ which {\em is} a well formed query, as `count' is an aggregate operation applicable on $b$-values. By the way, what this query returns is the number of branches which were delivered the same quantity of products. 

To see another example of analytic query, suppose that $T$ is a set of tweets accumulated over a year; $dd$ is the function associating each tweet $t$ with the date $dd(t)$ in which the tweet was published; and $cc$ is the function associating each tweet $t$ with its character count, $cc(t)$. To find the average number of characters in a tweet by date, we follow the same steps as in our running example: first, group the tweets by date (using function $dd$); then find the number of characters per tweet (using function $cc$); and finally take the average of the character counts in each group (using `average' as the aggregate operation). The appropriate query formulation in this case is the triple $(dd, cc, avg)$. 



Clearly the grouping and measuring functions in the above examples can be any functions, and in particular they can be the answers to traversal queries. This simple observation motivates the general definition of analytic query that we use in our model.

\begin{definition}
    Let $\mathcal C$ be a context. An {\em analytic query} over  $\mathcal C$ is a triple $(g, m, op)$, where $g$ and $m$ are tree queries  having the same source, and $op$ is an aggregate operation applicable over the target of $m$. Given a database $\delta$ over  $\mathcal C$, the answer of $Q$ over $\delta$, denoted $Ans(Q, \delta)$, or $Ans_Q$ when $\delta$ is understood,  is the following function:

    \smallskip\noindent $Ans_Q: range(g) \to target(op)$ such that: 
    
    \noindent $Ans_Q(i)= op(m/(g^{-1}(i)))$, for all $i$ in $range(g)$ \\

    \noindent Here $m/(g^{-1}(i))$ denotes the restriction of $m$ to the subset $g^{-1}(i)$ of its source. 
    
\end{definition}



The reason why the grouping and measuring queries $g$ and $m$ in the above definition are required to be tree queries is the following: grouping the objects in the common source of $g$ and $m$ means partitioning that source into disjoint blocks of objects; and unless the query $g$ returns a function we cannot obtain a partition. Similarly, unless the query $m$ returns a function there might be ambiguity in the measured property of an object.  By requiring that both $g$ and $m$ be tree queries, we ensure that their values are indeed functions. Clearly, the queries  $g$ and $m$ can also be traversal queries provided that each satisfy the equality constraint. 

A few remarks are in order here explaining in what ways the definition of analytic query given in this paper generalizes the one given in \cite{SpyratosS18}, significantly. First, by allowing grouping and measuring to be done by the results of traversal queries, we gain access to a much larger set of meaningful functions for grouping and measuring (`meaningful' in the sense that they are defined through queries expressing user needs). Second, we can benefit from the combination of the rewriting rules for traversal queries with the rewriting rules for analytic queries (that we shall see in the following section) so as to obtain a richer rewriting system for analytic queries. 

It is important to note that, in an analytic query, the common source of the grouping and measuring query can be any node - not necessarily the root of the context. For example, in Figure~\ref{CxtDef}(b), the common source of the grouping and measuring query of the following two analytic queries is $Prod$:  

\item $(s, c, count)$ returning the number of product categories supplied by each supplier   

\item $(c, s, count)$ returning the number of suppliers supplying each product category   \\

Clearly, given a context $\mathcal C$, we can use the identity edge $\iota_X$ and the terminal edge $\tau_X$ of each node $X$ in the same way as any other edge of $\mathcal C$. In particular, we can use them to form traversal queries and we can use such queries in defining analytic queries. Referring to the context of Figure~\ref{CxtDef}(b), here are two examples of analytic queries using the special edges, $\iota_A$ and $\tau_A$: 

\begin{center} 
$Q_1= (\iota_{Inv}, q, sum)$ and $Q_2= (q, \iota_{Inv}, count)$
\end{center}
\noindent During the evaluation of $Q_1$, in the grouping step, the function $\iota_{Inv}$ puts each invoice of $Inv$ in a  singleton block. Therefore summing up the values of $q$ in a block simply finds the value of $q$ on the single invoice in that block; then the measuring step simply returns this value of $q$. It follows that $Ans_{Q_1} = q$.

As for the query $Q_2$, the grouping query $q$ groups together all invoices having the same delivered quantity; and as $\iota_{Inv}$ doesn't change the values in each block, the answer to $Q_2$ is the number of invoices by quantity 
delivered. Here are two more examples: 

\begin{itemize}
    \item $(\tau_{Inv}, \iota_{Inv}, count)$ returns the number of all invoices in node $Inv$    \\
Note that the identity function $\iota_A$ is typically used for finding the cardinality of a node $A$ 

\item $(\tau_{Inv}, q, sum)$ returns the total of all quantities delivered (i.e. for all dates, branches and products)\\
Note that the constant function $\tau_A$ is typically used for finding the reduction of the whole of $A$ under some 
measuring query.  
\end{itemize}

It is important to note that combinations of queries such as the above can be used in several special cases such as when we want to compute percentages. For example, in Figure~\ref{CxtDef}(b), suppose we want to compute, for each branch, the percentage  of the total quantity of  products delivered. In this case, we use the following two queries: 

$Q_{Branch}= (b, q, sum)$, returning the total quantity delivered by branch

$Q_{Tot}= (\tau_{Inv}, q, sum)$, returning the total of all quantities delivered 

\noindent By dividing now the answer $Ans_{Q_{Branch}}(i)$ by the answer  $Ans_{Q_{Tot}}$ we find the desired percentage for each branch $i$.\\


A last remark regarding analytic queries as defined in this paper is that we have several possibilities of restriction. Here are a few examples:  

\begin{enumerate}
    \item We can restrict one or more nodes from those appearing in the grouping and measuring queries, as usual.   \\ 

    \item We can restrict the answer of the analytic query itself. Indeed, as the answer of an analytic query $Q$ is a function, we can restrict it to a subset $D$ of its domain of definition (or to a subset of its range and `push' the restriction  on its domain of definition). This kind of restriction is denoted as $Ans_Q/D$. For example, consider the query $Q= (p, q, sum)$ over the context of  Figure~~\ref{Price}, which returns the totals by product. Its answer is a function from $Prod$ to $Totals$, and if we define $D= \{x \in Prod / Ans_Q(x) \leq 1000\}$ then the restricted answer $Ans_Q/D$ will contain only products for which the total is less than or equal to 1000.   \\

    \item We can use any function which is applicable on the source of $Ans_Q$ to define quite complex restrictions of $Ans_Q$. For example, consider the query: $Q= (s, c, count)$ over the context of  Figure~\ref{Price}, which returns the number of categories supplied by supplier. Its answer is a function from $Sup$ to $CountCat$, and if we define $D= \{x\in Sup / Ans_Q(x) > Avg(Ans_Q(Sup))\}$ then the restricted answer $Ans_Q/D$ will contain only suppliers supplying more categories than the average of categories supplied by a supplier. 
\end{enumerate} 
We note that the kinds of restriction described in items 2 and 3 above are not possible in relational algebra queries (although possible in SQL group-by queries through the `Having' clause). So the question is: does our model offer all possibilities of restriction offered by the `Having' clause of SQL? The answer is yes, and a detailed account of this claim can be found in \cite{LS-submitted}. 

\subsection{Evaluation of analytic queries }\label{sec:EAQ}

The evaluation of an analytic query can be done in three different ways: (a) directly, based on the definition of analytic query, (b) by translating the query and evaluating it by some query engine, and (c) by rewriting the query so as to put it in some desired form. We discuss these three possibilities below.\\

\noindent {\em Direct evaluation} \\

\noindent Direct evaluation consists essentially in programming the steps described in the example seen in the previous subsection (see also Figure \ref{Eval}). Clearly, direct evaluation might prove inefficient for analysing big data sets.\\

\noindent {\em Translation} \\

\noindent Translating the query and evaluating it by some query engine takes advantage of optimization techniques implemented in the query engine. An analytic query as defined here can be translated as an SQL group-by query, when processing relational data \cite{SpyratosS18}; as a MapReduce job, when processing data residing in a file system \cite{SpyratosS18,ZervoudakisKSP21}; and as a SPARQL query when processing RDF data \cite{PapadakiST21}.  Moreover, the possibility of translating an analytic query as a query in three different kinds of query engines makes it possible to use the context model as a `mediator' \cite{DBLP:journals/csur/Wiederhold95}. This means that a user formulates an analytic query over the context; the query is translated to a query over the underlying query engine; and finally the user receives the answer `transparently', that is, as if the query were processed by the context. We have implemented such a mediator as described in the following section.\\

\noindent {\em Rewriting} \\ 

\noindent Rewriting an analytic query can be done at two orthogonal levels: (a) rewriting the grouping and/or measuring query, and (b) rewriting the analytic query itself in terms of other analytic queries. Rewriting the grouping or the measuring query can be done as explained in section \ref{TQEval}. Rewriting the analytic query in terms of other analytic queries depends on the form of the grouping and measuring queries as well as on the kind of aggregate operation used in the analytic query. More precisely, if the aggregate operation is associative \cite{LaurentS11}, then the following is the basic rule for rewriting an analytic query \cite{SpyratosS18}: \\

{\bf {\em Composition Rule}} :  $(g' \circ g, m, op)= (g', (g, m, op), op)$ \\

\smallskip\noindent What this rule says is that if the grouping query is the composition of two other queries then the analytic query can be rewritten as a nested query. To understand this nesting, consider the following analytic query over the context of Figure \ref{CxtDef}(b): \\
$Q= (r \circ b, q, sum)$, asking for the totals by region \\
\noindent Here is why the composition rule works: the edge $r$ tells us in which region each branch is located, therefore, if we have the totals for each branch in a region then we can sum them up to find the total for that region. Now, to find the totals by branch we can use the analytic query $Q'= (b, m, q)$; and as the answer to $Q'$ has the same source as $r$ we can use the analytic query  $Q''= (r, Ans_{Q'}, sum)$ to compute the totals by region. In other words, the query $Q$ for computing the totals by region can be rewritten as $Q= (r, Ans_{Q'}, sum)$. By replacing now $Ans_{Q'}$ by $(b, m, op)$ we obtain that  $Q= (r, (b, q, sum), q)$, which should be read as follows: first evaluate the inner or nested query $(b, m, op)$ to obtain the totals by branch; and then use the answer as the measure to evaluate the outer query to obtain the totals by region. 

It is important to note that the composition rule works only if the aggregate operation $op$ is associative \cite{LaurentS11}. Most common operations such as `sum', `min', `max' etc. are associative. For example, $sum(1, 2, 3, 4, 5)=sum(sum(1, 2), sum(3, 4, 5))$, whereas `avg' is not. For example, $avg(1, 2, 3, 4, 5) \neq avg(avg(1, 2), avg(3, 4, 5))$). Although there are `corrective' algorithms allowing the use of many non-associative operations such as average, median, etc. (see for example~ \cite{ZervoudakisKSP21}\cite{LaurentS11}), we shall not pursue this subject any further here. Rather, in order to simplify the discussion, we shall tacitly assume that all aggregate operations are associative.

\begin{figure}
{
\begin{center}
\includegraphics[width=350px,keepaspectratio]{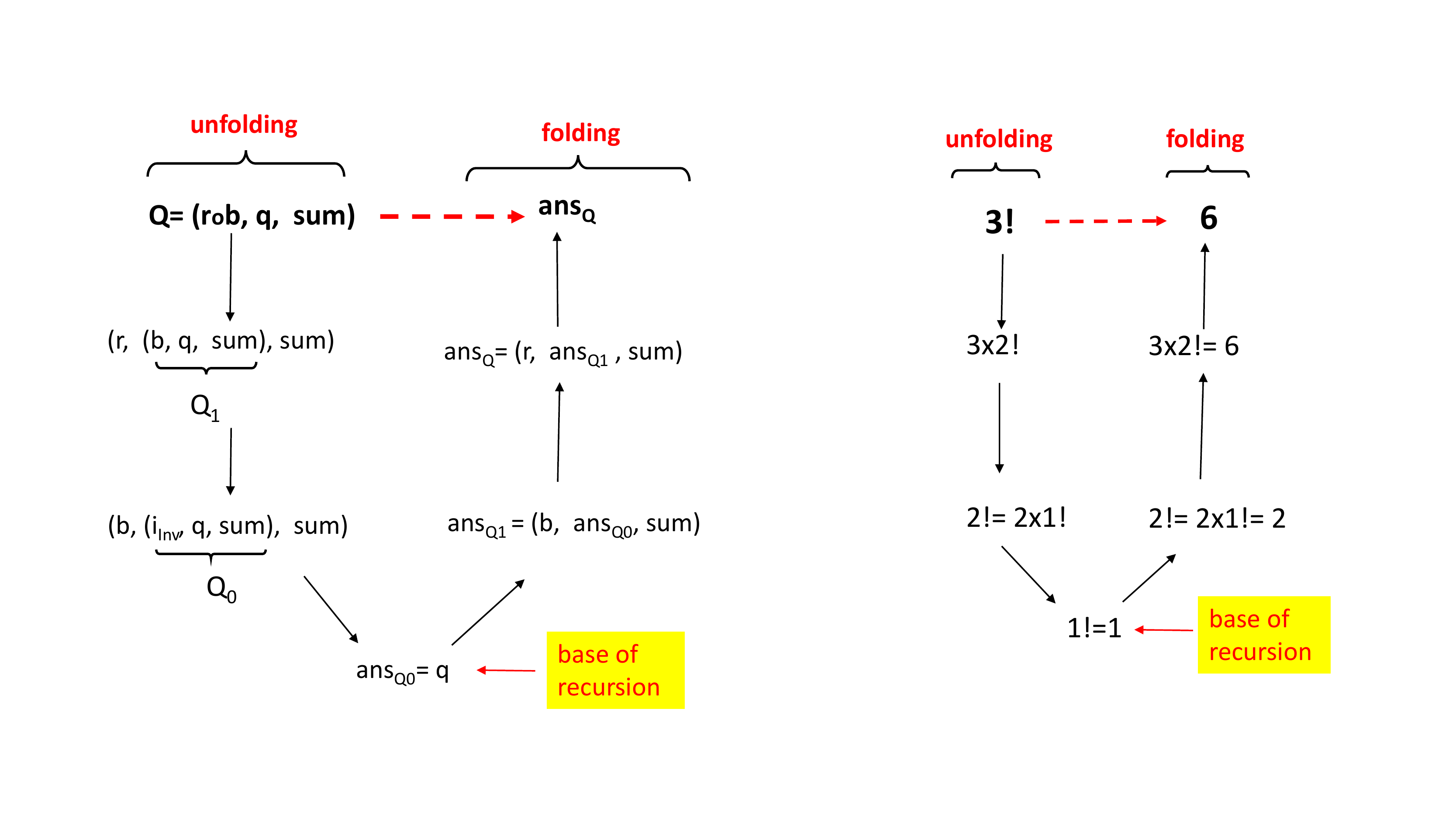}
\caption{Recursive evaluation of an analytic query\label{Cxt9}}
\end{center}
}
\end{figure}

The composition rule allows for the recursive evaluation of analytic queries as shown on the left of Figure \ref{Cxt9}, where we use as an example the query $Q= (r \circ b, q, sum)$, asking for totals by region. We first unfold $Q$ using the composition rule until we reach the base query $Q_0$; then we evaluate the base query and use its answer to compute the answer to the query $Q_1= (b, q, sum)$; and finally we use the answer of $Q_1$ to compute the answer of $Q$. Note how the identity function $\iota_{Inv}$ is indispensable in defining the base query $Q_0$, based on the equality: $b= b \circ \iota_{Inv}$. Indeed, it follows from this equality that: 

~~$Q_1= (b, q, sum)= (b \circ \iota_{Inv}, q, sum)= (b, (\iota_{Inv}, q, sum), sum)$ 

~~Therefore $Q_0= (\iota_{Inv}, q, sum)$ and $Ans_{Q_0}= q$ 

\noindent Also note the similarity between this recursive evaluation of $Q$ and the recursive evaluation of $3!$, shown on the right of Figure \ref{Cxt9}. 

Using the composition rule  we can sometimes avoid making joins during the evaluation of an analytic query. To see how, consider again the query $Q= (r \circ b, q, sum)$ and suppose that $r$ and $b$ are stored on separate files, say $R$ and $B$, respectively. The direct evaluation of $Q$ requires a join between $R$ and $B$ in order to find the grouping function, whereas if we use the rewritten form $Q'= (r, (b, q, sum), sum)$, no join is needed since the grouping functions for the inner and outer query are stored in $R$ and $B$ respectively. Another advantage of using the composition rule is the possibility of incremental evaluation of query answers when processing big data sets \cite{SpyratosS18}\cite{ZervoudakisKSP21}. 

Now, the composition rule says how to rewrite an analytic query in terms of other analytic queries, no matter whether we rewrite or not its grouping and measuring queries. Let's now see an example of rewriting an analytic query at both levels: first rewriting its grouping and/or measuring query to obtain some desired (equivalent) form; then rewriting the resulting analytic query using the composition rule. For example, consider the following analytic query:

$((p \circ s) \wedge (p \circ c), q, sum)$  asking for the totals by supplier and category
 
\noindent Here, by applying lemma \ref{BasicLemma} to the grouping query $(p \circ s) \wedge (p \circ c)$ we can rewrite it as $(s \wedge c) \circ p$, so we can now rewrite the analytic query as follows:

$((s \wedge c) \circ p, q, sum)$

\noindent Next, we apply the composition rule to obtain the final, rewritten query:

$((s \wedge c), (p, q, sum), sum)$

\smallskip\noindent 
\noindent As this example shows, using the composition rule together with rewriting of the grouping and/or measuring query, we can derive a variety of rewriting rules. A most notable example is what we call the `pairing rule'. Consider an analytic query of the form $Q= (g_1 \wedge g_2, m, op)$, where the grouping query is the pairing of two queries, $g_1$ and $g_2$. Suppose now that we want to evaluate the query $Q_1=(g_1, m, op)$. It follows from Lemma  \ref{BasicLemma} that $g_1= \pi_{target(g_1)} \circ (g_1 \wedge g_2$, therefore $Q_1$ can be rewritten as follows:  ~$Q_1=(\pi_{target(g_1)} \circ (g_1 \wedge g_2), m, op)$ \\

\noindent Applying now the composition rule, we can further rewrite $Q_1$ as follows:\\
$Q_1= (\pi_{target(g_1)} \circ (g_1 \wedge g_2), (g_1 \wedge g_2, m, op), op)$\\

\noindent Following the same reasoning for the query $Q_2= (g_2, m, op)$ we have:\\
$Q_2= (g_2, m, op)= (\pi_{target(g_2)} \circ (g_1 \wedge g_2), (g_1 \wedge g_2, m, op), op)$ \\

\noindent Hence the following rule for rewriting a query $Q_i= (g_i, m, op)$ in terms of the query $Q= (g_1 \wedge \ldots \wedge g_n, m, op)$, $n \geq 2$ :  \\

\noindent {\bf {\em Pairing Rule}}:  \\
$Q_i= (g_i, m, op)= (\pi_{target(g_i)} \circ (g_1 \wedge \ldots  \wedge g_n), (g_1 \wedge \ldots  \wedge g_n, m, op), op)$,  $n \geq 2$ \\


\smallskip\noindent 
To see how the pairing rule works, refer to Figure \ref{CxtDef}(b) and suppose that we have computed the totals by branch and product. Then we can re-use this result to compute the totals by branch and the totals by product in terms of the totals by branch and product. Indeed, applying the pairing rule we have:

\noindent $(b, q, sum)= (\pi_{Branch} \circ (b \wedge p), (b \wedge p, q, sum), sum)$ 

\noindent $(p, q, sum)= (\pi_{Product}\circ (b \wedge p), (b \wedge p, q, sum), sum)$.\\

A last remark on analytic queries: analytic queries can be used in defining views in much the same way as traversal queries can (see section \ref{Views}). For example, in Figure \ref{CxtDef}(b), one can add the following edges as views over the context: \\
$\mathcal V_1$: $Date \times Branch \to TotQty$ with label $e_1$, where $e_1=(d \wedge b, q, sum)$\\
$\mathcal V_2$: $Branch \to TotQty$ with label $e_2$, where $e_2= (b, q, sum)$ \\
$\mathcal V_3$: $Region \to TotQty$ with label $e_3$, where $e_3= (r \circ b, q, sum)$\\
Such views may be useful, for example, for stock management or for product delivery planning in the enterprise.

\section{Applications}\label{sec:Apps} 

In this section we describe two applications of our model, namely: (a) how to define a relational database as a view of a context, and (b) how  to use a context as a user-friendly interface for analysing relational data.

\subsection{Defining a relational database over a context}\label{Embedding}

As we have seen in section \ref{TQDef} (item 3), each tree query $Q(K, E_1, \ldots, E_n)$ {\em induces} a relation schema $R_Q(K, A_1, \ldots, A_n)$,  where $A_1= target(E_1), \ldots, A_n= target(E_n)$; and for every database $\delta$, the answer $Ans_Q$ {\em induces} a relation $r_Q$ over $R_Q$. Moreover, from Lemma \ref{BasicLemma} we have that $\pi_{A_i}\circ Ans_Q$ is a function from $K$ to $A_i$, $i= 1, \ldots, n$, and therefore the relation $r_Q$ {\em satisfies} the functional dependencies $K \to A_1, \ldots, K \to A_n$. We exploit this relationship between tree queries and relation schemas in order to propose a method for defining relational databases over a context. 

First, let us recall this relationship between tree queries and relation schemas through a simple example. Consider the tree query $Q(Prod, s, c)$ over the context of Figure \ref{CxtDef}(b). Its key is $Prod$ and it has two expressions, $s$ and $c$, with targets $Sup$ and $Cat$, respectively. Therefore $Q$ induces a relation schema $R_Q(Prod, Sup, Cat)$; and for every database $\delta$, its answer is $Ans_Q= \delta(s) \wedge \delta(p)$ {\em inducing} a relation $r_Q$ (i.e. a set of triples) defined as follows: 

$r_Q= \{(x, y, z)\} / x \in \delta(Prod), y= \delta(s)(x), z= \delta(c)(x)\}$

\noindent Therefore $Q$ induces a relation schema $R_Q(Prod, Sup, Cat)$ with key $K$, and a relation $r_Q= Ans_Q$ over $R_Q$ satisfying the dependencies $Prod \to Sup$ and $Prod \to Cat$ since $\delta(s): \delta(Prod) \to \delta(Sup)$ and $\delta(c):\delta(Prod) \to \delta(Cat)$ are functions.

Note that two different tree queries may induce the same relation schema as in the following example. 

\smallskip
$Q(Inv, r \circ b, c \circ p)$ 

$Q'(Inv, h \circ s \circ p, c \circ p)$. 

\smallskip\noindent Indeed, both $Q$ and $Q'$ induce the relation schema
$R(Inv, Region, Cat)$ with dependencies $Inv \to Region$ and $Inv \to Cat$. Note however that the attribute $Region$ has different semantics in the two relations, as its semantics in $Q$ is provided by the expression $r \circ b$ whereas in $Q'$ its semantics is provided by the expression $h \circ s \circ p$. 

\noindent Clearly, having two or more different tree queries inducing the same relation schema $R$ presents no problem. On the contrary, it provides the flexibility of choosing the semantics we want to associate with $R$.

\smallskip
Summarizing our discussion so far, one can define a relational database schema over a context $\mathcal C$ by simply defining and storing a set of tree queries over $\mathcal C$ (i.e. by defining a view over $\mathcal C$). Moreover, each relation over that database schema will be consistent with respect to its key dependencies. 

It is important to note that, in defining a relational database as a view of a context, we have a significant semantic gain: a relation schema $R_Q(K, A_1, \ldots, A_n)$ defined by a tree query $Q$, comes with a clear semantics for its attributes, this semantics being the expression defining each attribute (as in our previous example of queries $Q$ and $Q'$). In contrast, in a relational database, the attributes of a relation schema have only the intuitive semantics conveyed by their names. Therefore when a relational database is defined as a view of a context, the underlying context works as the `semantic layer' of the defined database. 

Now, what we have seen so far using tree queries can be extended to traversal queries. The main problem here is that an attribute may be the target of two or more parallel expressions. We can solve this problem using one of two approaches: either require that the traversal query satisfy all equalities (see section \ref{IC}) or use `aliases'. To illustrate these two approaches consider the following example of traversal query: 

$Q(Inv, h \circ s \circ p, r \circ b, c \circ p)$

\noindent The first two expressions of $Q$, namely  $E_1= h \circ s \circ p$ and  $E_2= r \circ b, c \circ p$, are parallel expressions with $Region$ as their common target. Following the first approach, we can declare: 

$Q(Inv, E_1, E_2)$ with $E_1= E_2$ 

\noindent As a result of this declaration, the answer of $Q$ becomes a tree and therefore the induced relation schema is $R_Q(Inv, Region)$, and for every database the induced relation $r_Q$ satisfies the key dependency $Inv \to Region$. 

\noindent If we follow the second approach, we have to provide as many `aliases' for each attribute $A$ as there are expressions in the query with $A$ as their common target (obviously, no alias is needed if $A$ is the target of a single expression). We shall use the following convention: if $E_1, \ldots, E_n$, $n \geq 2$, are the parallel expressions of $Q$ with $A$ as their common target then replace $A$ in the induced relation schema by the following $n$ attribute names: $E_1.A, \ldots, E_n.A$ with $dom(E_1.A)= \ldots = dom( E_n.A)= dom(A)$. Let us illustrate this point further using our previous example: 

$Q(Inv, h \circ s \circ p, r \circ b, c \circ p)$ 

 \noindent The induced relation schema is: 
 
$R_Q(E_1.Region, E_2.Region, Cat)$ 

\noindent Assuming this aliasing convention, we can define the relation schema induced by a traversal query as in the case of a tree query. In practice, one can use more suggestive aliases, such as $Sup.Region$ and $Branch.$ $Region$ in our example, provided that each alias is associated with a single expression. Note that such aliases are called `attribute renamings' in relational databases. Clearly, for a tree query, no such aliases are needed. Finally, the choice of one between the two approaches described above depends on the application envisaged. \\

\noindent We end this subsection by two remarks concerning relational databases defined as views over a context:\\
\begin{enumerate}
    \item One can define more than one database over the same context. Clearly, when two or more such databases co-exist, querying poses no problem: each database can be queried independently of the others. However, updating requires caution as, even if the databases are virtual, updating one database may affect the contents of the others. More generally, as databases are views, one encounters all problems related to view management \cite{DBLP:journals/debu/GuptaM95}.
    \item The user of a database defined as a view of a context (through a set of traversal queries) actually `sees' a set of relation schemas (i.e. a set of tables), exactly as when interacting with a traditional relational database. Here, however, in contrast to a traditional relational database, each table is associated with clear semantics that the user can inspect. Indeed, if the attributes are made clickable then the  interested user can click an attribute of a table and `see' the expression used to compute the values of that attribute. This expression says how the functions of the underlying context are combined in order to compute the value of that attribute. Moreover, it is natural that the user of such a database can ask relational algebra queries, as usual, therefore the system will have to translate users' queries as traversal queries over the underlying context. Although tedious, this translation is indeed possible to make. We shall not go into the details of the translation in this paper. The interested reader is referred to \cite{LS-submitted} for a complete account.
\end{enumerate} 

\noindent Summarizing our discussion in this subsection we have seen that: (a) given a context $\mathcal C$, we can define a consistent relational database as a view of the context, that is, as a set of traversal queries and (b) each relation of such a database has clear semantics provided by the expressions of the defining query.

\subsection{A context as an interface for data analytics}\label{subsec:Interface} 

\begin{figure}
{
\begin{center}
\includegraphics[width=350px,keepaspectratio]{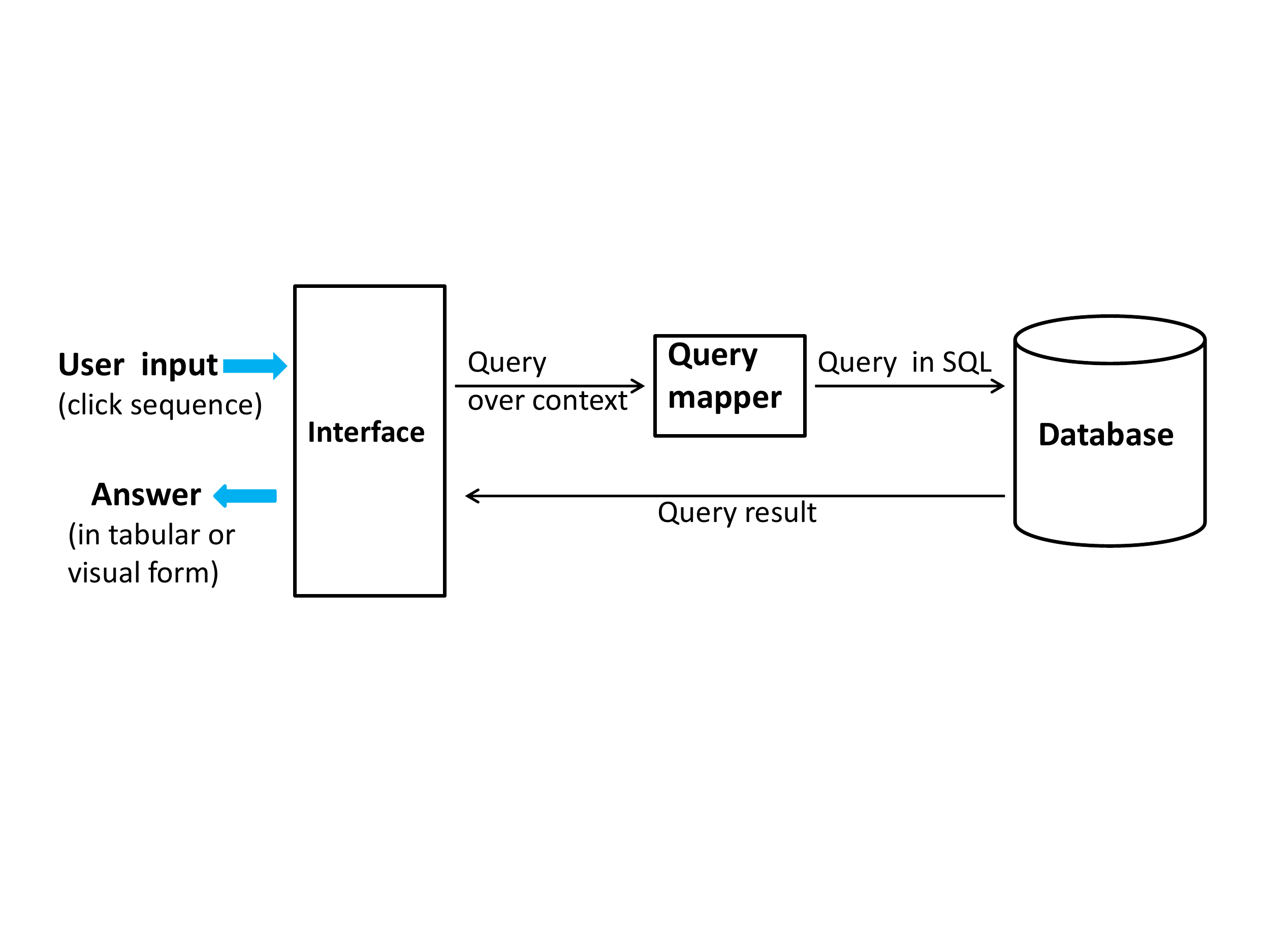}
\caption{Information flow in the interface\label{Flow}}
\end{center}
}
\end{figure}

In this section we show how a context can serve as a user-friendly interface to a relational database for data analysis purposes. The general idea works as follows: (a) represent the database by a context as explained in the introductory section (see Figure~\ref{CxtDef}), and make its nodes clickable; (b) the user of the context defines an analytic query through a sequence of clicks that the interface translates as an SQL group-by query on the underlying database; and (c) the user receives the result in the form of a table and/or in a visual form following some visualization template. Our idea is depicted in the diagram of Figure~\ref{Flow} that we will elaborate further shortly.

In the interface that we have designed and implemented \cite{KaterinaNMS2022}, interaction between the user and the interface occurs in four steps  as follows:

\smallskip
\noindent {\em Step 0}: The relational database, call it RDB, is represented as a context, say RDB-Context (as we saw in the introduction) and the interface shows to the user this context with its nodes made clickable.

\noindent {\em Step 1}: The user clicks on a set of nodes $A_1$, \ldots, $A_n$ of the RDB-context (meaning that the user requests if there is a relation over these nodes). 

\noindent {\em Step 2}: The system responds by showing to the user $n$ proposals. Each proposal consists of  a set $E(K, A_i)$ of parallel expressions with source $K$ and target $A_i$, $i= 1, \ldots, n$. 

\noindent {\em Step 3}: The user selects one proposal and one expression $E_i$ from $E(K, A_i)$, for each $A_i$, $i= 1, \ldots, n$ (and eventually defines restrictions on the nodes of the selected expressions). This defines the {\em user's analysis context}, call it $\mathcal C_u$, which is a tree. The user can now ask analytic queries over $\mathcal C_u$ through a sequence of clicks (as will be described shortly). 

\noindent {\em Step 4}: The interface translates each analytic query submitted by the user as an SQL group-by query $Q$ on the underlying relational database. The query is then evaluated and its result is returned to the user in the form of a table and/or in some other visual form. \smallskip

The information flow in the interface is described succinctly by the diagram of Figure~\ref{Flow}. Additionally, the interface provides a zooming facility when the context graph is very large, so that the user can concentrate on a sub-graph of interest of (conceptually) manageable size, before starting with Step 1. 



Note that once a relation has been extracted the interface can also be used to define usual relational algebra queries. For example, to define a projection of the extracted relation it is sufficient to select the keyword `projection' (from a menu) and then click on the attributes over which projection is to be done; to define a selection it is sufficient to select the keyword `selection' and then click on attributes, one by one, giving the value(s) to be selected for each of the clicked attributes; and so on. For more details the user is referred to \cite{KaterinaNMS2022}. 

To illustrate the interaction between the user and the system consider a data warehouse containing the relations $R, R1, R2, R3, R4$ as defined in the introduction. Setting up the interface requires two preliminary actions: (a) define the RDB-context which, in this case, is the context shown in Figure~\ref{Price} as explained in the introduction, and (b) select the visualization templates to be used in the interface. These actions constitute the preliminary Step 0.

The information flow in the interface follows the diagram of Figure~\ref{Flow}, where the user's clicks are translated by a mapper into an SQL group-by query on the underlying relational database; then the query is evaluated and the result is sent to the interface. As an option, the user can click a desired visualization template from a pop-up menu so as to visualize the result according to the selected template. In order to test the interface, we have experimented with the Pentaho-Mondrian food database (http://mondrian.pentaho.com) as described in detail in \cite{KaterinaNMS2022}. 

To illustrate how an analytic query is defined by the user through a sequence of clicks, consider again the context of Figure \ref{Price}. A user who wants to define the analytic query `totals by branch's $Region$' (formally: $(r \circ b, q, sum)$) will go through the following steps:\\

\noindent {\em Grouping mode}: The user clicks on the nodes $Inv, Branch, Region$ thus defining the expression $r \circ b$ 

\noindent {\em Measuring mode}: The user clicks on nodes $Inv, Qty$ thus defining the expression $q$

\noindent At this point the interface determines the aggregate operations applicable on $Qty$ and shows to the user a pop-up menu containing these operations; the user clicks one (or more) operations in the menu. 

\noindent {\em Result visualization mode}: The user clicks a visualization template from a pop-up menu (tabular, pie, scatter plot, histogram etc.) \\

\noindent The above actions specify completely the analytic query, as `clicked' by the user, as well as the form in which the user will receive the result. \\

\noindent We are currently incorporating in the interface the rewriting rules that we have seen for traversal queries, as well as for analytic queries.

\section{Concluding remarks and perspectives}\label{sec: Conclusions}

We have seen a novel graph database model that we call the context model, as well as its query language consisting of two kinds of queries, traversal queries and analytic queries. The definition of both these kinds of queries is based on a simple functional algebra. We have also presented two non trivial applications of the model that demonstrate its expressive power, namely the possibility to define a consistent relational database as a view of a context, and also the possibility to use a context as a user-friendly interface for analysing relational data.

Our current work lies in the area of relational databases and tries to exploit the relationship between traversal queries and relation schemas in order to revisit some basic problems of relational database theory. Our approach can be succinctly described as follows: Let $U= \{A_1, \ldots, A_n\}$ be a set of attributes and $F$ a set of functional dependencies over $U$ such that every attribute of $U$ appears on the left or on the right side of at least one dependency of $F$. Seen as a set of functions, $F$ induces an acyclic graph $G_F$ over $U$ up to relation schema equivalence, where schema equivalence is defined as follows:  $X \equiv Y$ if the dependencies $X \to Y$ and $Y \to X$ are both implied by $F$, for all $X, Y \subseteq U$. As  $G_F$ is acyclic it may have one or more roots, therefore, in general,  $G_F$ will consist of one or more contexts (eventually sharing nodes and/or edges as in Figure \ref{TwoRoots}. We use this framework to revisit two basic aspects of database theory ~\cite{Ullman}: (a) the theory of functional dependencies and (b) the classical problem of designing a `good' relational database schema {\em and}  the semantics of its tables implied by $F$.



Our future work follows two main research lines. First, we would like to define update and transaction languages for the context model. Updating in a context $\mathcal C$ can happen at two levels: updating the graph $\mathcal C$ or updating the database $\delta$.  Updating the graph $\mathcal C$ means adding or removing edges under the constraint that the graph remain acyclic. In a relational database, this operation corresponds to changing the database schema which is extremely complex and costly as it means migrating the data of the database to the new schema (a problem related to data exchange \cite{FKMP03,FKP05}). On the other hand, updating the database $\delta$ means inserting, removing or modifying a pair $(x, f(x))$, where $f$ is a function of $\delta$. As such updates must preserve function totality, they may provoke updates in other functions of $\delta$, eventually on all functions of $\delta$. Therefore the need for efficient algorithms for the propagation of updates throughout the database. 

The second line of research concerns the extension of the context model. As it stands now, the context model is a model for data management and data analytics. We would like to extend it so as to provide support also for information visualisation and visual exploration. To this end we plan to extend the functional algebra with additional operations along the lines of \cite{SpyratosS19}.

\section*{Acknowledgements}
The author would like to thank Professor Dominique Laurent for his comments that helped improve the content of this paper.


\bibliographystyle{plain}
\bibliography{biblio}

\end{document}